\documentclass[a4paper,11pt]{article}
\usepackage{jinstpub}
\usepackage{wasysym}
\usepackage{comment}
\usepackage{lineno}
\usepackage{color}
\usepackage{anyfontsize}

\title{A nonlinear multiphysics model for the design validation of the ASTAROTH copper-steel cryogenic chamber.}

\author[1]{F.~Alessandria,}
\author[2,1]{F.B.~Armani,}
\author[1]{S.~Coelli,}
\author[3]{D.~Cortis,}
\author[2,1]{D.~D'Angelo,}
\author[1]{E.~Martinenghi,}
\author[1]{M.~Monti,}
\author[3]{D.~Orlandi,}
\author[2,1]{M.~Sorbi,}
\author[2,1]{V.~Toso,}
\author[1]{A.~Zani}

\affiliation[1]{INFN Milano, via G. Celoria 16, 20133 Milano (MI), Italy}
\affiliation[2]{Dipartimento di Fisica, Universit\`a degli Studi di Milano, via G. Celoria 16, 20133 Milano (MI), Italy}
\affiliation[3]{INFN LNGS, via G. Acitelli 22, 67100 Assergi (AQ), Italy}

\emailAdd{andrea.zani@mi.infn.it}

\abstract{
Among the global efforts to directly detect dark matter, the only positive claim so far relies on NaI(Tl) crystal detectors, making this technology of particular interest. ASTAROTH is a project aimed at developing the next generation of such detectors by reading out their scintillation light with SiPM matrices operated at cryogenic temperatures.
This paper describes the innovative design of the ASTAROTH cryostat, consisting of a double-walled copper-steel cryogenic chamber that cools the detectors by means of a liquid argon bath. The detectors are thermalized in a helium atmosphere at a temperature tunable from 87 to 150 K.
The design has been validated in terms of heat transfer efficiency and mechanical stress, developing a nonlinear multiphysics model.
The mechanical properties of OFHC copper were experimentally evaluated on dedicated tensile samples.
The simulation results show that the structural integrity is guaranteed. At the highest operating temperature, the region with the steepest temperature gradient exhibits stresses that slightly exceed the yield strength of copper (localized strain-hardened condition).
Following construction, the cryostat was commissioned and has been in regular operation for over 30 cooling cycles, with no signs of degradation. The temperature can be tuned across the full target range and remains stable within 0.1 K. These results demonstrate that this is a viable design for next-generation dark matter detectors, as well as for a variety of applications requiring uniform and tunable gas-conducted cooling of instrumentation.
}


\begin{document}
\maketitle
\flushbottom

\section{Introduction}
\label{sec:intro}

\subsection{The case for physics}
\label{sec:physics}

Several gravitational phenomena at galactic and extra-galactic scales~\cite{Sofue2001, Clowe2006, Freese2009} suggest the presence of a significant amount of unseen mass.
The overall picture points to about 85\% of mass in the Universe that is still undetected. Referred to as dark matter, it could differ in nature from standard baryonic matter, and it might be made of one or more particles yet to be discovered~\cite{Bertone2010}. The Standard Halo Model~\cite{Asgari2023halo} of the Milky Way galaxy pictures the visible counterpart of our galaxy as embedded in a quasi-spherical halo of dark matter. In this hypothesis, Earth would move through a ``wind'' of particles, which could be detected upon interacting with a target.
In fact, a (very) weak interaction with ordinary matter (other than gravitational) is foreseen by most theoretical models~\cite{cirelli2024}, thus leading to a world-wide experimental effort to detect it.
To protect them from cosmic radiation, dark matter detectors are usually placed in underground locations; additionally, they are specifically designed to have very low content of radioisotopes in the target and constructing materials, and are actively and/or passively shielded from intrinsic and environmental radioactivity. Given the very low expected energy depositions and event rates of these interactions, as well as the large uncertainties on the nature and properties of dark matter particles, these detectors are at the frontier of technology.

Though experiments with different technologies have been conducted in the past 30~years~\cite{damic, CRESST, Agnes2018_2, PandaX2025, SuperCDMS}, the only experimental observation to date that could be reconciled with the interaction of dark matter comes from the DAMA experiment~\cite{Bernabei2021}. DAMA was hosted at the underground Gran Sasso National Laboratory (LNGS) and used as target an array of NaI(Tl) scintillating crystals. The extraordinary claim requires to be confirmed by an independent measurement using the same target material and a few attempts to this aim are indeed ongoing~\cite{anais2025, cosine2025, Calaprice2022}. 

Technologically, DAMA and most similar experiments share a detector design where several few-kg NaI(Tl) crystals (cylindrical or similarly shaped) are coupled to photomultiplier tubes (PMTs). NaI(Tl) is a high-performance scintillator, emitting $40-50$~photons per keV of deposited energy (ph/keV)~\cite{Sasaki2006}. The emission peaks at around~420~nm, and it is usually collected by two PMTs placed on two faces of the crystal, whereas its remaining surface is wrapped in reflective material. 
This technology, however, has shown limitations under a few aspects, the main being the high number of noise events from the PMTs. These exceed the number of scintillation events by up to two orders of magnitude at the few~keV energies of interest, limiting the sensitivity to detect dark matter interactions~\cite{Calaprice2022}.

\subsection{The ASTAROTH project}
\label{sec:astaroth}

The ASTAROTH project~\cite{Dangelo2023} aims to overcome the described limitations, by adopting arrays of silicon photomultipliers (SiPMs) for the readout of scintillation light from NaI(Tl) crystals. 
SiPMs have several advantages over PMTs in terms of lower noise, compactness, limited radioactivity, higher light conversion efficiency~\cite{Dinu2016}. A detector based on SiPMs must operate at cryogenic temperatures, due to their high dark count rate at room temperature, which renders them unsuitable for low energy applications. On the other hand, the latest generation of SiPMs shows a dark count rate up to two orders of magnitude lower than typical PMTs for the same active surface, when operated below 100~K~\cite{Acerbi2017}. When devising a low-temperature environment for SiPMs, liquid argon (LAr) comes up as a natural solution, as it is a good scintillator~\cite{Heindl_2010} and can be instrumented as a secondary veto detector surrounding the crystals. This last feature is strategic for the environmental/material radioactive background suppression~\cite{Antonello2019}. One difficulty of this configuration lies in the fact that the light response of NaI(Tl) crystals at low temperature is not well known: the few available measurements in literature ~\cite{Sibczynski2011, Sailer2012} point to the fact that the response is not linear with temperature and it might actually be heavily dependent on the specific manufacture recipe of the producer (e.g., ingot production technique, fraction of thallium). Furthermore, direct immersion of crystals in LAr should be avoided, favoring a slow and controlled cool-down that prevents the development of cracks. For all these reasons, the experiment requires a flexible system that allows stable operation of the crystals in a gaseous atmosphere, in a temperature ranging from 87~K (LAr bath) to about 150~K. The latter is the approximate temperature above which the SiPMs start to exhibit a dark count rate higher than PMTs. Such a system would allow to effectively characterize the crystals at different temperatures, and therefore identify the best operational point of the detector (crystal+SiPMs) for physics runs.

This paper presents the design validation of a novel double-walled copper-steel cryogenic chamber. Developed within the ASTAROTH project, it enables the slow and controlled cooling of NaI(Tl) scintillating crystals by thermal coupling to a surrounding LAr bath. Although designed for this specific application, the chamber has broader utility and has already been used for the characterization of materials, electronic components, circuit boards, and integrated circuits over a broad temperature range, while avoiding the thermal conduction issues that often affect similar measurements when performed with cold heads.

\section{Double-walled chamber description}
\label{sec:chamber}

\begin{figure}
\centering
\includegraphics[width=.8\textwidth]{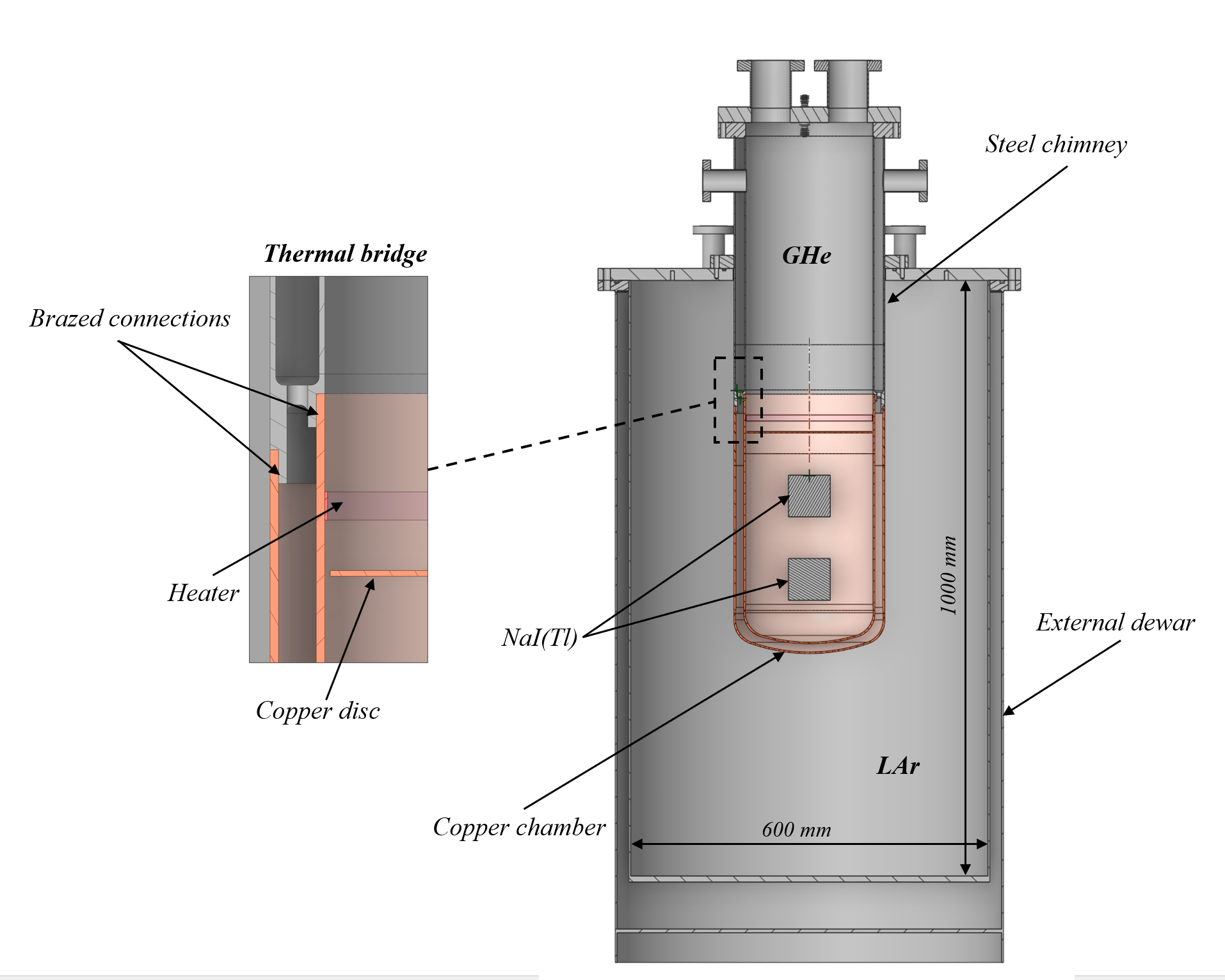}
\caption{ASTAROTH cryostat: the double-walled copper-steel cryogenic chamber and chimney is installed in the outer dewar and hosts two detectors. Only the lowest of five anti-convection discs in the chimney is shown.\label{fig:chamber}}
\end{figure}

\subsection{Design requirements}
\label{sec:design}

The cryogenic chamber was designed to safely and uniformly cool up to two cubic NaI(Tl) crystals within the 87–150~K temperature range. 
The crystals have linear dimensions of 50~mm and are protected from atmospheric moisture by a fully transparent gas-tight case (see also Section \ref{sec:commissioning}). They can be instrumented with SiPM arrays on up to all six surfaces. 

The temperature gradients that the crystals can tolerate during cool-down are specified by the manufacturer: (1) the temporal gradient must be limited to less than 20~K/h; (2) the difference between any two points must not exceed 1~K.  
Before designing the chamber, these conditions were verified during the cool-down test of a prototype crystal in a dedicated setup.  
A third design requirement (3) regards the temperature stability during data taking, which must be maintained within 0.1~K to ensure a steady response from both the crystal and the SiPMs.

The need to reliably thermalize the crystals and SiPMs to a stable, uniform, and well-known temperature requires a gas-filled environment. For this reason, cold head cryostats, operating in vacuum, were considered unsuitable for this purpose.

\subsection{Technical description}
\label{sec:technical}

The main body of the chamber is made by two concentrical cylinders of oxygen-free high conductivity (OFHC) copper (Figure~\ref{fig:chamber}). 
At the bottom, each cylinder is closed by a torospherical endcap\footnote{This shape was chosen during the design phase, based on simulations performed by the Mechanical Service of INFN Milano \cite{milano1}, to minimize stress caused by evacuation of the inter-wall cavity.}.  
At the top, they are connected to a chimney composed of a second pair of concentric cylinders.  
These are made of AISI 316 L stainless steel (SS); they have the same radial dimensions as the chamber cylinders, and terminate at the top with a flanged connection.  
In this way, the chamber and chimney form a single, double-walled, vacuum-insulated volume.  
The outer OFHC copper wall has a diameter of 253~mm, with a cavity width of 13.5~mm. Considering each wall is 3~mm thick, the inner clearance measures 214~mm.

The copper-steel connection is realized by brazing the two elements on both the inner and outer walls. Just above the brazed connection, the two walls of the chimney terminate in a so-called 'thermal bridge' (see Figure~\ref{fig:chamber}). This element provides a preferential pathway for heat conduction between the outer cooling bath (Section~\ref{sec:bath}) and the inner volume housing the detectors (Section~\ref{sec:heat}).  

Sixteen holes of diameter 5.5~mm are drilled in the thermal bridge to allow the chamber and chimney inter-wall cavities to form a single evacuation volume, which can be pumped through a port located on the side of the chimney at room temperature.

The chamber and chimney are 463.5~mm and 428~mm high, respectively, yielding a total inner volume height of 891.5~mm.  
The chimney features two side Ultra High Vacuum (UHV) CF DN35 penetrations: one, as mentioned, is used to pump the inter-wall volume; the other provides access to the main volume and is generally used for reading temperature sensors.  
The chimney is closed by a top flange that hosts gas connections for helium circulation, a safety valve, and a few UHV sub-flanges. These allow for significant flexibility in routing detector connections.

The chamber and chimney system is connected to the main flange of a cylindrical dewar (diameter 600~mm, height 1000~mm), and extends 247~mm above the dewar flange.  In this way, the OFHC copper chamber can be completely immersed in the cryogen filling the dewar volume, leaving a safe ullage of around 50~mm. 

The design of the chamber was initially developed at INFN Milano, through internal design reviews and preliminary mechanical simulations performed with the support of the Mechanical and Design Service \cite{milano1}. The final, executive design, in particular concerning the geometry of the thermal bridge, the required surfaces for brazing, and the number/diameter of the holes in the bridge, was developed together with the production company.

\subsection{Cryogenic bath}
\label{sec:bath}

The design of the chamber exploits the natural cooling power provided by an external cryogenic bath.
As mentioned in Section \ref{sec:astaroth}, liquid argon (boiling point 87~K) is the designed cryogen for the physics runs and has been assumed in the design validation presented in this paper.
However, for commissioning and initial operations (Section~\ref{sec:commissioning}), liquid nitrogen (LN$_2$, boiling point 77~K) was used for convenience. The same approach was adopted for determining the cryogenic properties of the materials (Section~\ref{sec:materials}). The 10~K difference in boiling point between the two cryogens is considered small enough that the cryostat operation is not significantly affected.

The cryogenic bath is hosted in the external dewar, whose main flange is equipped with UHV ports for cryogen injection and extraction as well as feedthroughs for level sensors. The maximal operating pressure of the dewar is 0.5~barg. A cryogenic safety relief valve, dimensioned specifically for the present use case, guarantees this pressure is not exceeded.

\subsection{Heat transfer and control}
\label{sec:heat}

By design, heat transfer between the inner volume and the cryogenic bath occurs primarily through conduction across the thermal bridge, ensuring the slow cooling of the detectors.
The radiative heat exchange between the two walls of the chamber has been computed to be 60-100\footnote{The range depends on the value of emissivity, $\epsilon$, assumed for OFHC copper, which can have a significant variability in literature.} times lower than conduction through the bridge, and therefore considered negligible. 
The top of the chimney is exposed to ambient temperature; while heat conduction through its thin walls is negligible, gas convection may still compromise the thermal equilibrium of the chamber. Therefore, five thin disks are installed along the chimney height to promote gas stratification and limit convective cycles. The disk closest to the crystals, located at the thermal bridge height, is made of OFHC copper and is the only one included in the model. The other four are made of AISI 316 L and do not appear in Figure~\ref{fig:chamber}.

The crystals are suspended at the center of the chamber by a support attached to four non-conductive polymeric rods anchored to the inner side of the top chimney flange.

As mentioned, the thermal contact between the chamber and the crystal is done by filling the inner volume with gas.  
Dry helium is chosen to avoid condensation at operating temperatures. The target pressure is 100~mbar, where heat transfer is optimal~\cite{he-props}.

An adjustable 150~W resistive heater is installed on the inner surface of the chamber, immediately below the steel-copper transition. The heater power is regulated via an external power supply, and it allows to raise the chamber temperature in a controlled manner up to any target temperature between the cryogen boiling temperature and 150~K, and to maintain it stable.

\subsection{Materials and construction choices}
\label{sec:choices}

The rationale for adopting a two-material design is as follows. OFHC copper is preferred for the chamber, as it is the material closest to the detectors and can be sourced with very low (radio)contamination. 
Its high heat conductivity ensures swift uniform distribution of the heat throughout the inner wall of the chamber. 
AISI 316 L, on the other hand, is preferred for the chimney and bridge due to its mechanical strength and compatibility with UHV components, which are made of the same material. 
Concerning thermal behavior, the chimney is exposed to room temperature for the part above the flange of the dewar, while the bridge is designed to regulate the cooling power entering the main chamber, preventing a rapid cool-down that could damage the crystals. Also in this respect AISI 316 L was preferable for both elements due to its higher thermal impedance with respect to OFHC copper (Figure ~\ref{fig:ct}).
The copper-steel connection is achieved by high vacuum brazing the two elements. This process ensures uniform thermal treatment of the materials during construction.

At the highest operating temperatures, a temperature difference of up to about 70~K exists across the thermal bridge and between the two chamber walls. Consequently the mechanical stresses on the bridge and on the copper-steel junction have been analyzed with particular attention in the structural simulation described in Section~\ref{sec:simulation_structural}.

\section{Material properties and cryogenic characterization}
\label{sec:materials}

In this section we describe the thermal and mechanical properties of the materials (AISI 316 L and OFHC copper) of the chamber that have been assumed or measured in view of the thermomechanical simulation described in Section~\ref{sec:simulation}.

\subsection{Thermal properties}
\label{sec:materials_thermal}

\begin{figure}
\centering
\includegraphics[width=.85\textwidth]{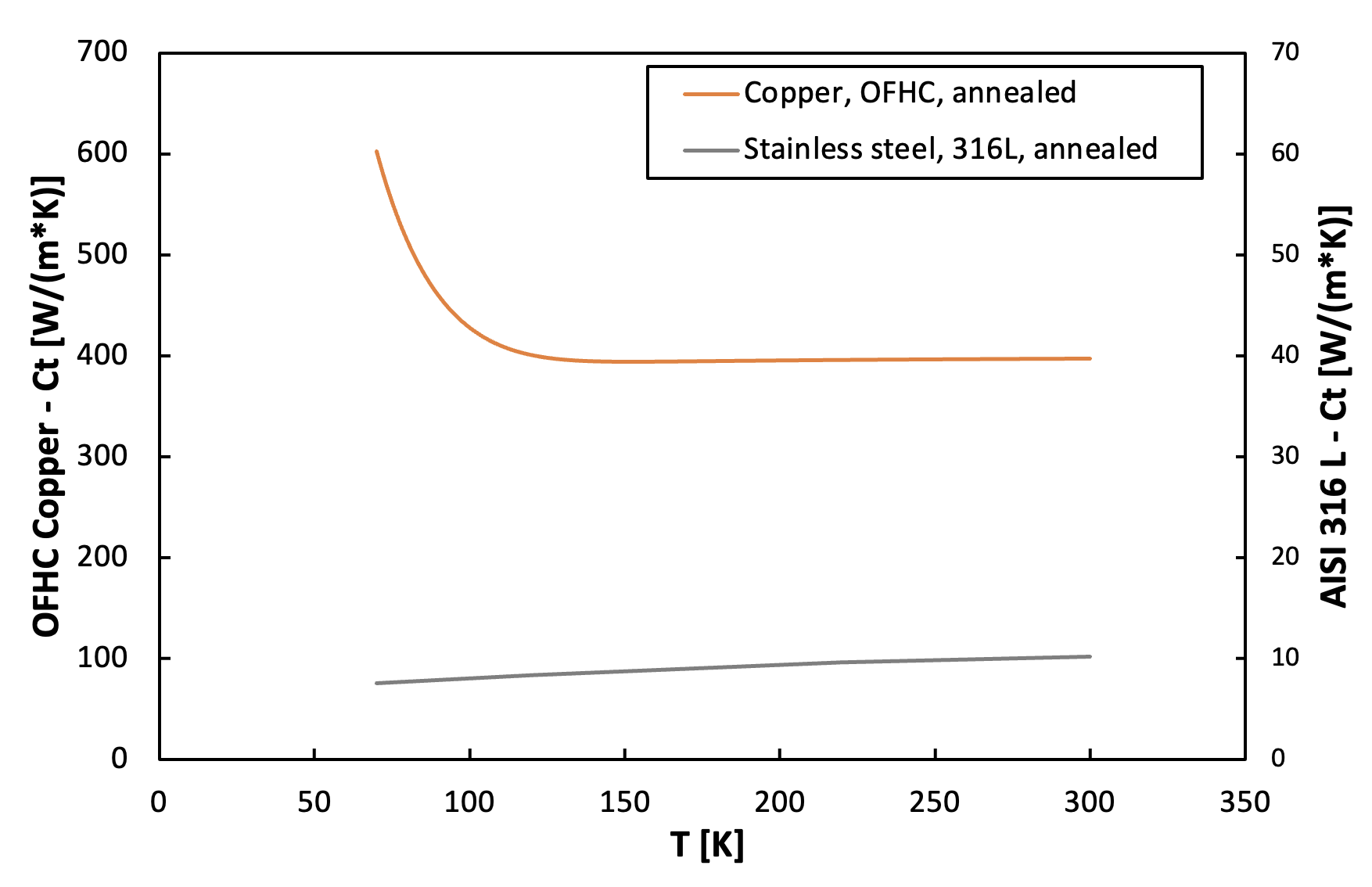}
\caption{Thermal conductivity ($C_t$) of OFHC copper and AISI 316 L~\cite{copper, steel}. \label{fig:ct}}
\end{figure}

\begin{figure}
\centering
\includegraphics[width=.85\textwidth]{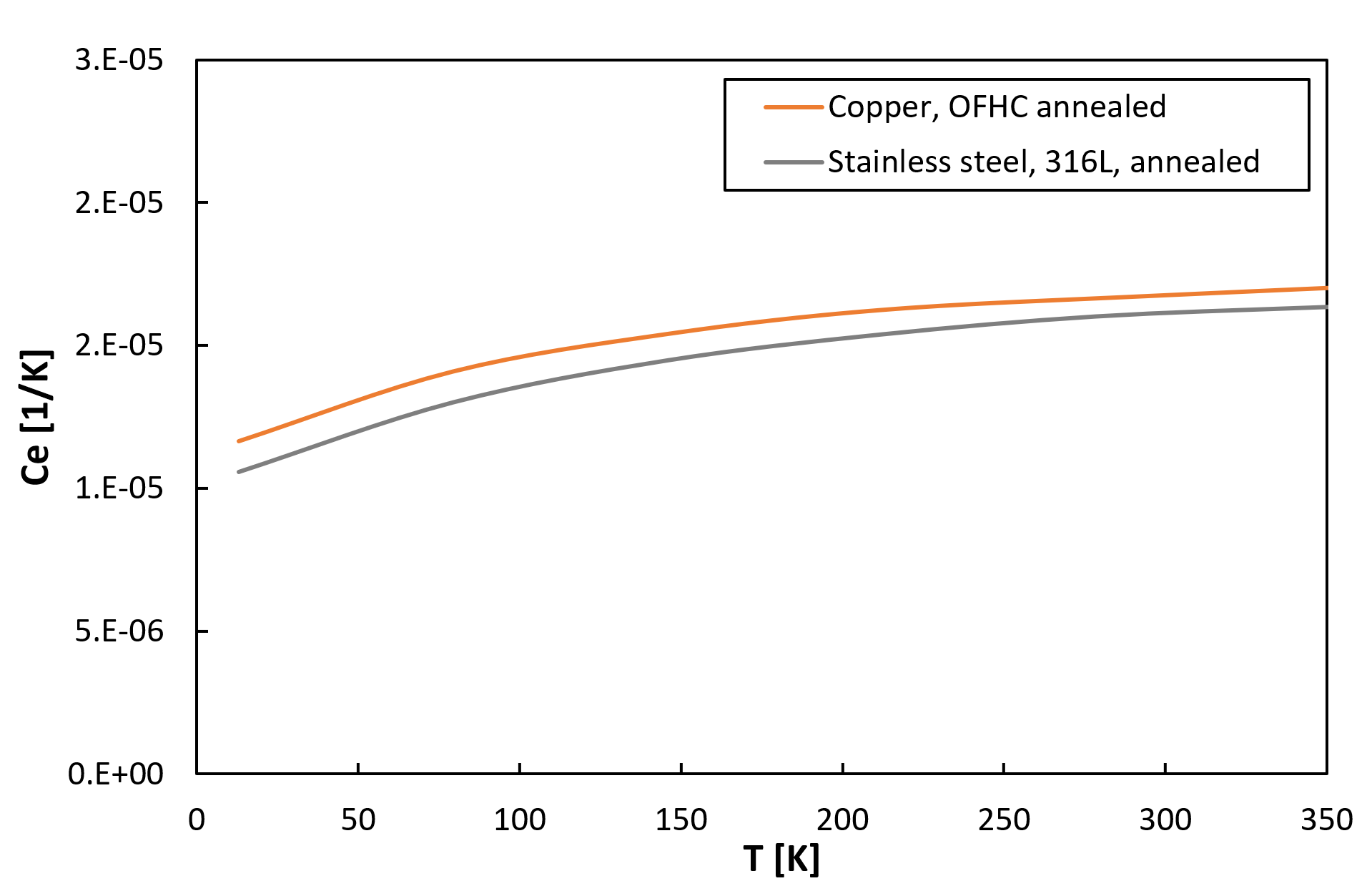}
\caption{Expansion coefficient ($C_e$) of OFHC copper and AISI 316~L~\cite{copper, steel}.  \label{fig:ce}}
\end{figure}

The thermal properties of OFHC copper and AISI 316 L down to cryogenic conditions were identified through a review of the literature and by means of the available databases~\cite{copper, steel, TOBLER} within the software used for the analysis: ANSYS Workbench \cite{ansys}. The behavior of the thermal conductivity ($C_t$) and the thermal expansion coefficient ($C_e$), used in the thermal simulations described in Section~\ref{sec:simulation_thermal}, are shown in Figures~\ref{fig:ct} and~\ref{fig:ce}. The similarity between the $C_e$ coefficients minimizes thermal stresses, while the difference in $C_t$ justifies the choices of these two materials for the different parts of the design, as explained in Section~\ref{sec:choices}.

\subsection{Mechanical properties}
\label{sec:materials_mechanical}

\begin{table}
    \centering

    \begin{tabular}{|c|c|c|c|c|} \hline 
         T [K]&  E [GPa]&  YS [MPa]& UTS [MPa] \\ \hline 
         295&  195&  216& 530
 \\ \hline 
         77&  209&  314& 1235
 \\ \hline
    \end{tabular}
    \caption{Theoretical mechanical properties of AISI 316~L~\cite{TOBLER}.\label{tab:316-mech}}
\end{table}

The mechanical properties of AISI 316 L are found to be mostly independent of the material shape and of the working process. Consequently, data available in the literature were used (Table~\ref{tab:316-mech}), in analogy to what was done for the thermal properties~\cite{TOBLER}.

\begin{figure}
\centering
\includegraphics[width=.8\textwidth]{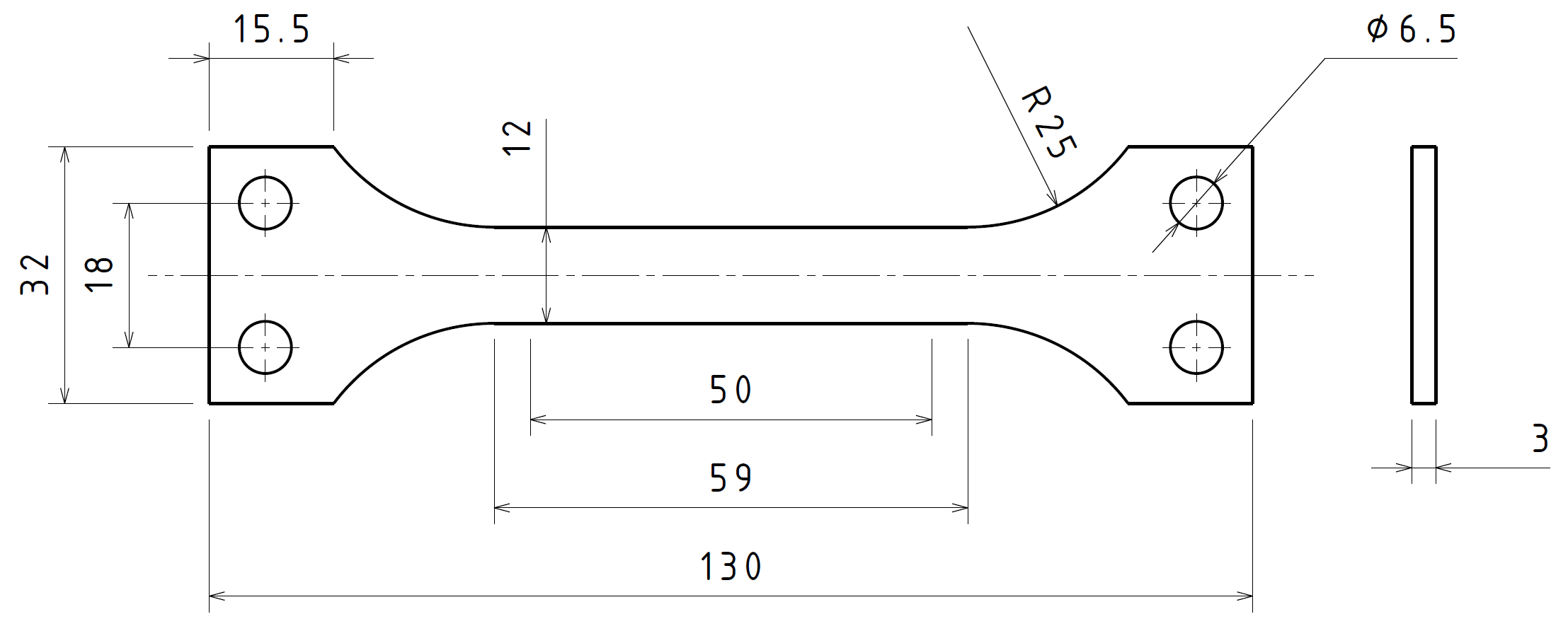}
\caption{Tensile test specimen geometry (mm). \label{fig:sample}}
\end{figure}

On the other hand, OFHC copper properties show a wide range of variability among the batches that can be found on the market. For this reason, it was decided to cut a few material samples from the production batch of the chamber and test them to evaluate the mechanical properties of the specific copper in use. 
The shape and dimensions of the tensile specimens are shown in Figure~\ref{fig:sample}. 

The specimens were annealed for 5 h up to about 850 °C prior to any test, reproducing the heating cycle that the material undergoes during construction due to the copper-steel brazing in oven. 

The tensile tests for the determination of the Elastic Modulus (E), Yield Strength (YS), Ultimate Tensile Strength (UTS), and Elongation (A), were conducted using a universal INSTRON testing machine equipped with a LN$_2$ cryostat, available at Accelerators and Applied Superconductivity Laboratory (LASA), in Segrate, Milan. Tests were carried out at room temperature and at 77~K (i.e., LN$_2$ temperature), after a slow cooling rate ($0.6-0.7$~K/min) and a proper stabilization. Three different specimens were tested at each temperature.

For each sample, subsequent loading and unloading cycles were performed, reaching a total strain of about 0.01 (1\%). During this phase, strain was measured using a custom strain gauge with a 10~mm gauge length and 1~mm strain range (10\%).
These cycles were intended to simulate the work hardening the material experiences during initial chamber operations, such as the first vacuum pump-down and initial cool-downs to operating conditions. 

\begin{figure}
\centering
\includegraphics[width=.70\textwidth]{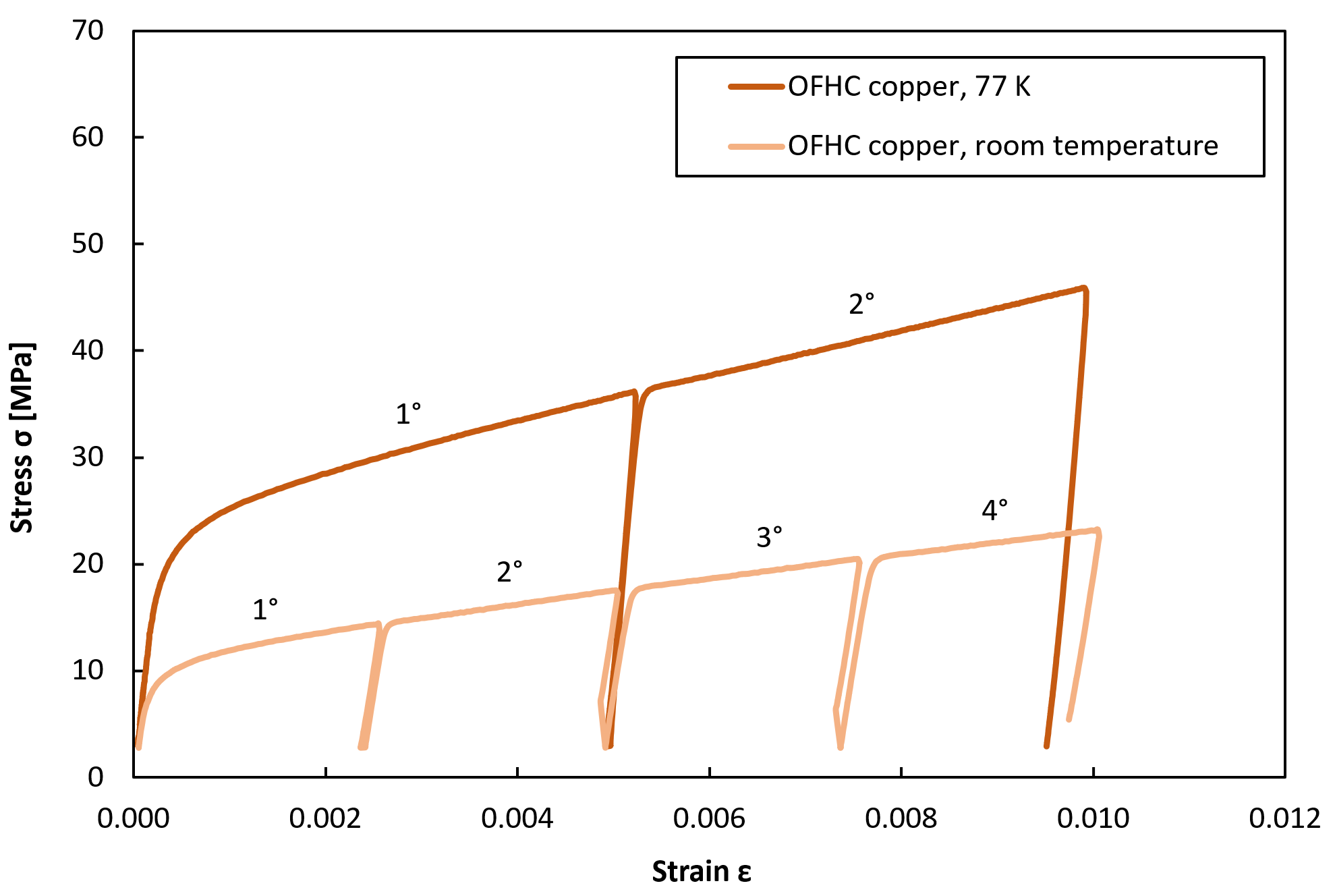}
\caption{Cyclic work hardening of OFHC copper up to 0.01 strain (1\%) at room temperature and 77~K. \label{fig:hardening}}
\end{figure}

Figure~\ref{fig:hardening} shows the loading and unloading cycles performed on two specimens, at room temperature and 77~K. 
It is possible to see that the plastic regions of the curves appear to belong to a single trend line. This behavior is expected and indicates the progressive increase of the elastic limit of the material (i.e., YS limit). 

\begin{figure}
\centering
\includegraphics[width=.70\textwidth]{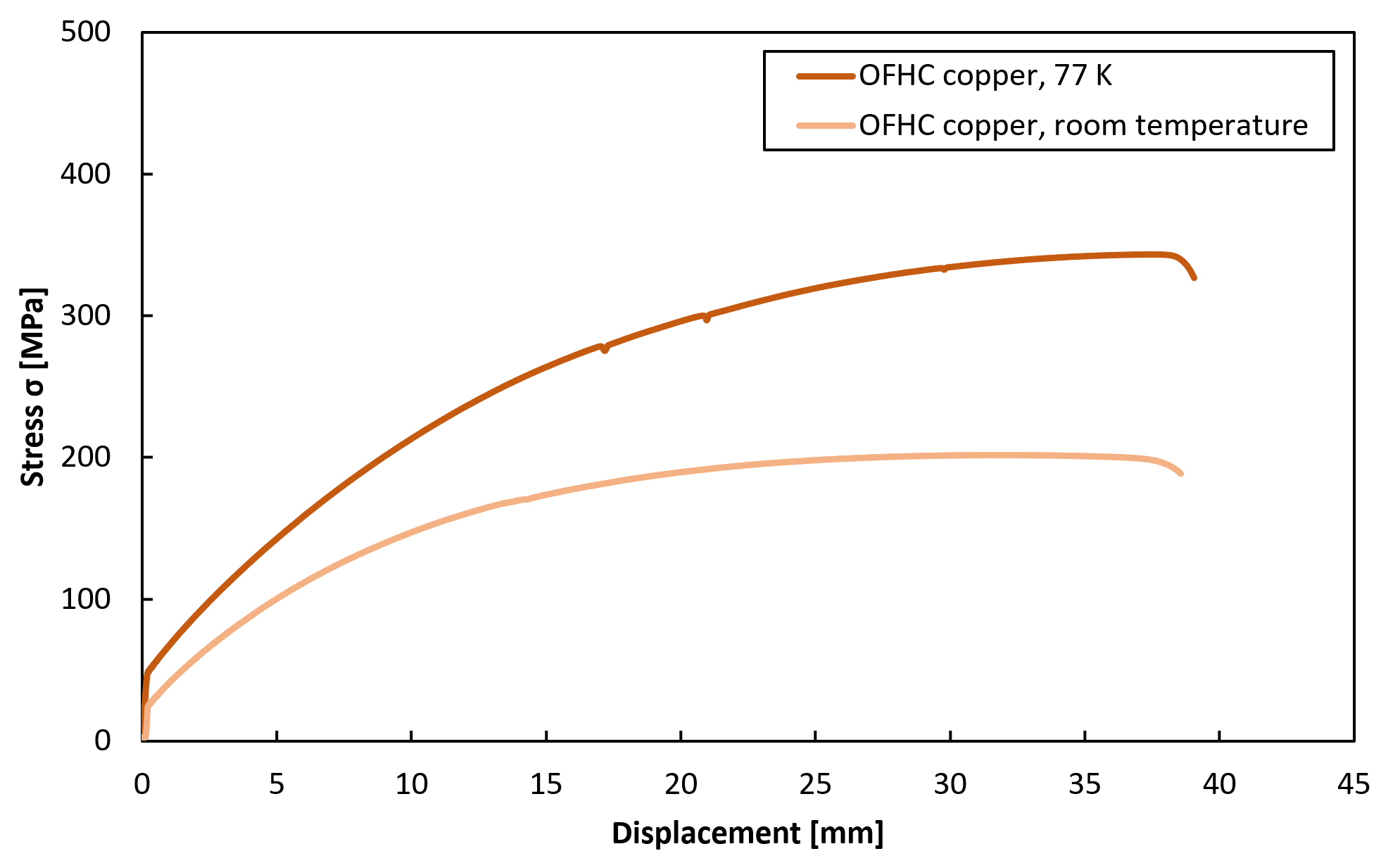}
\caption{Stress-displacement curve of OFHC copper up to rupture at room temperature and 77~K. \label{fig:stress-displacement}}
\end{figure}

After these hardening cycles, a tensile test up to rupture was done. This test exceeded the range of the strain gauge, therefore the deformation of the sample was measured directly as the displacement of the crossbar of the universal testing machine. 
Figure~\ref{fig:stress-displacement} shows the rupture tests for two specimens at room temperature and 77~K.

Higher YS and UTS values are observed for the cryogenic specimens compared to those at room temperature, which is consistent with literature findings (see~\cite{Cu-props} and references therein). 
However, rupture occurs at similar percentage elongation values ($\sim$~50\%) in both cases, demonstrating that the material retains its ability to absorb deformation energy when cooled.

The mechanical properties of the OFHC copper, reported as the mean values of the results obtained from the three specimens at each temperature, are shown in Table~\ref{tab:copper-mech}.

\begin{table}
    \centering

    \begin{tabular}{|c|c|c|c|c|} \hline 
         T [K]&  E [GPa]&  YS [MPa]& UTS [MPa] &A [\%]\\ \hline 
         295&  56.4 ± 0.8&  23.4 ± 1.0& 201 ± 3.5& 55.9 ± 0.8\\ \hline 
         77&  108.8 ± 0.3&  43.9 ± 6.2& 333.1 ± 2.1&48.5 ± 0.3\\ \hline
    \end{tabular}
    \caption{Experimental mechanical properties of OFHC copper. Elastic Modulus (E), Yield Strength (YS), Ultimate Tensile Strength (UTS), and Elongation (A). Mean values and 2$\sigma$ std deviations of the three measurements are reported.\label{tab:copper-mech}}
\end{table}

\begin{figure}
\centering
\includegraphics[width=.70\textwidth]{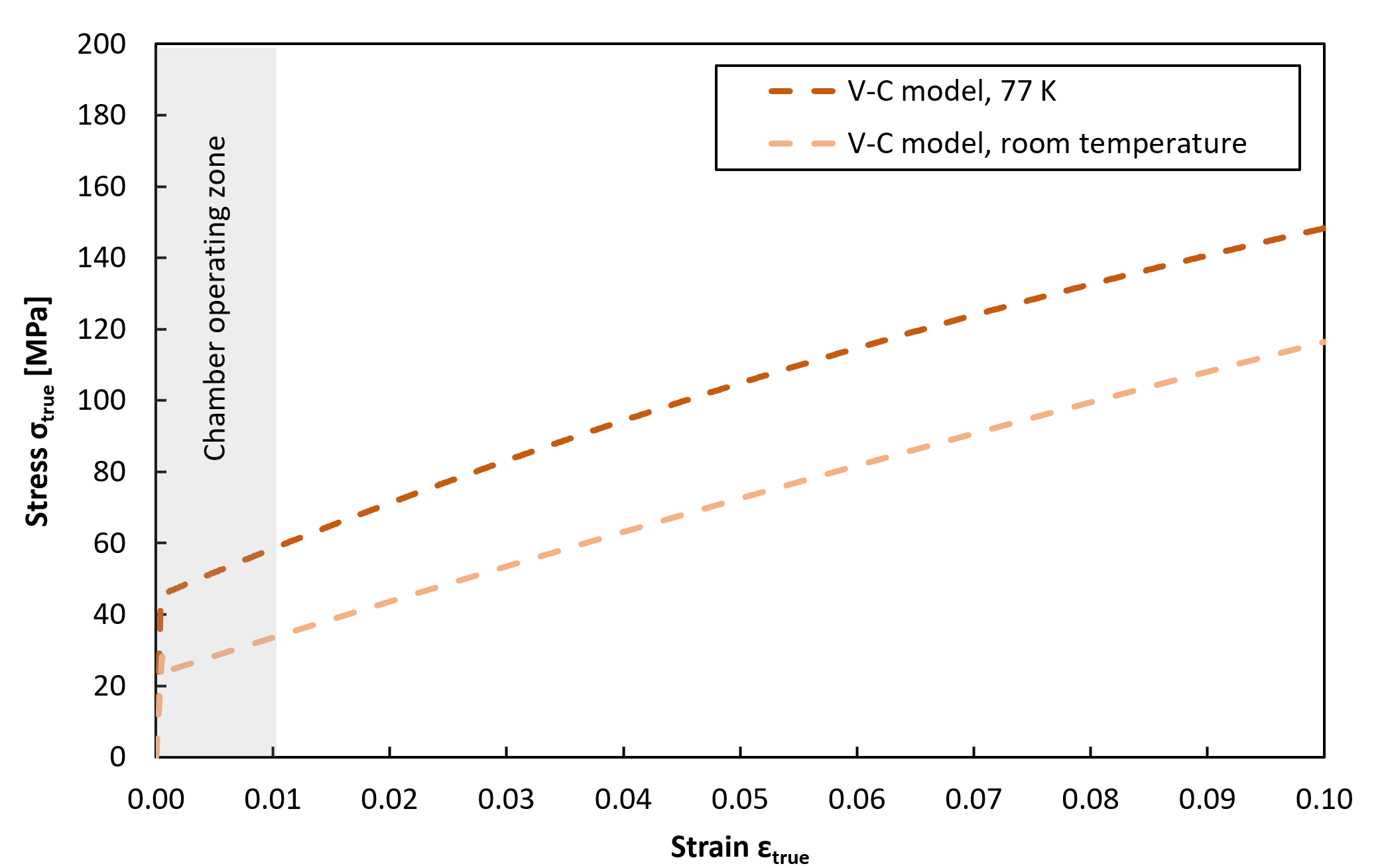}
\caption{V-C models of OFHC copper at room temperature and 77~K. 
\label{fig:vc}}
\end{figure}

In order to proceed with the simulation of the chamber (Section~\ref{sec:simulation_structural}), data
from the loading/unloading cycles have been used to calibrate an elasto-plastic model for the two tested temperatures. 
Data have been expressed in terms of true-strain $\varepsilon^{\text{t}} = \ln(1 + \varepsilon)$ and true-stress, 
$\sigma^{\text{t}} = \sigma (1 + \varepsilon)$.
The model that was chosen is the Voce-Chaboche (V-C)~\cite{petry}, since it is generally used in common finite element software, such as ANSYS Workbench. It is based on a negative exponential law describing the progressive decrease of true-stress with the increase of the plastic true-strain $\varepsilon^{\text{t}}_p$:
\begin{equation}
\sigma^{\text{t}}_{\text{model}} = c_1 + c_2 \left(
1 - e^{-c_3 \varepsilon^{\text{t}}_{\text{p}} } \right) + c_4 \varepsilon^{\text{t}}_{\text{p}}
\label{eq:v_c}
\end{equation}
Here, $c_1$ represents the yield strength (YS), $c_2$ the saturation coefficient, $c_3$ the exponential rate of stress variation, and $c_4$ models linear hardening or softening during plastic deformation (and can be zero).
The coefficients were determined using a nonlinear generalized reduced gradient (GRG) optimizer~\cite{grg}. 
The algorithm calculates at each iteration the gradient of the error function with respect to the experimental data:
\begin{equation}
r = \sum_{k=1}^{3} \sqrt{ \frac{ \sum_{j=1}^{N} \left( \sigma^{\text{t}}_{\text{measured}} - \sigma^{\text{t}}_{\text{model}} \right)^2 }{N} }
\label{eq:err}
\end{equation}
where $j$ indexes the $N$ data points within each specimen test and $k$ indexes the three tests.
The coefficients are reported in Table~\ref{tab:vc}, whereas the model stress-strain plot is shown in Figure~\ref{fig:vc}. 

Due to material hardening, a small plastic strain is expected to occur during the chamber commissioning.
The V-C models are expected to accurately describe all chamber deformations, including both elastic and plastic regions.

\begin{table}
    \centering

\begin{tabular}{|c|c|c|c|c|c|} \hline
        & c$_1$ & c$_2$ & c$_3$ & c$_4$ & r \\ \hline
295 K   & 25.3 & 443.1 & 2.4 & 0.0 & 0.5 \\ \hline
77 K    & 47.8 & 214.7 & 6.4 & 0.0 & 0.6 \\ \hline
\end{tabular}
\caption{OFHC copper at room temperature and 77~K: Voce-Chaboche (V-C) coefficients and the gradient of the error function. See equations~\ref{eq:v_c} and~\ref{eq:err} and text for definition.\label{tab:vc}}
\end{table}

\subsection{Test of the brazed connection}

\begin{figure}
\centering
\includegraphics[width=\textwidth]{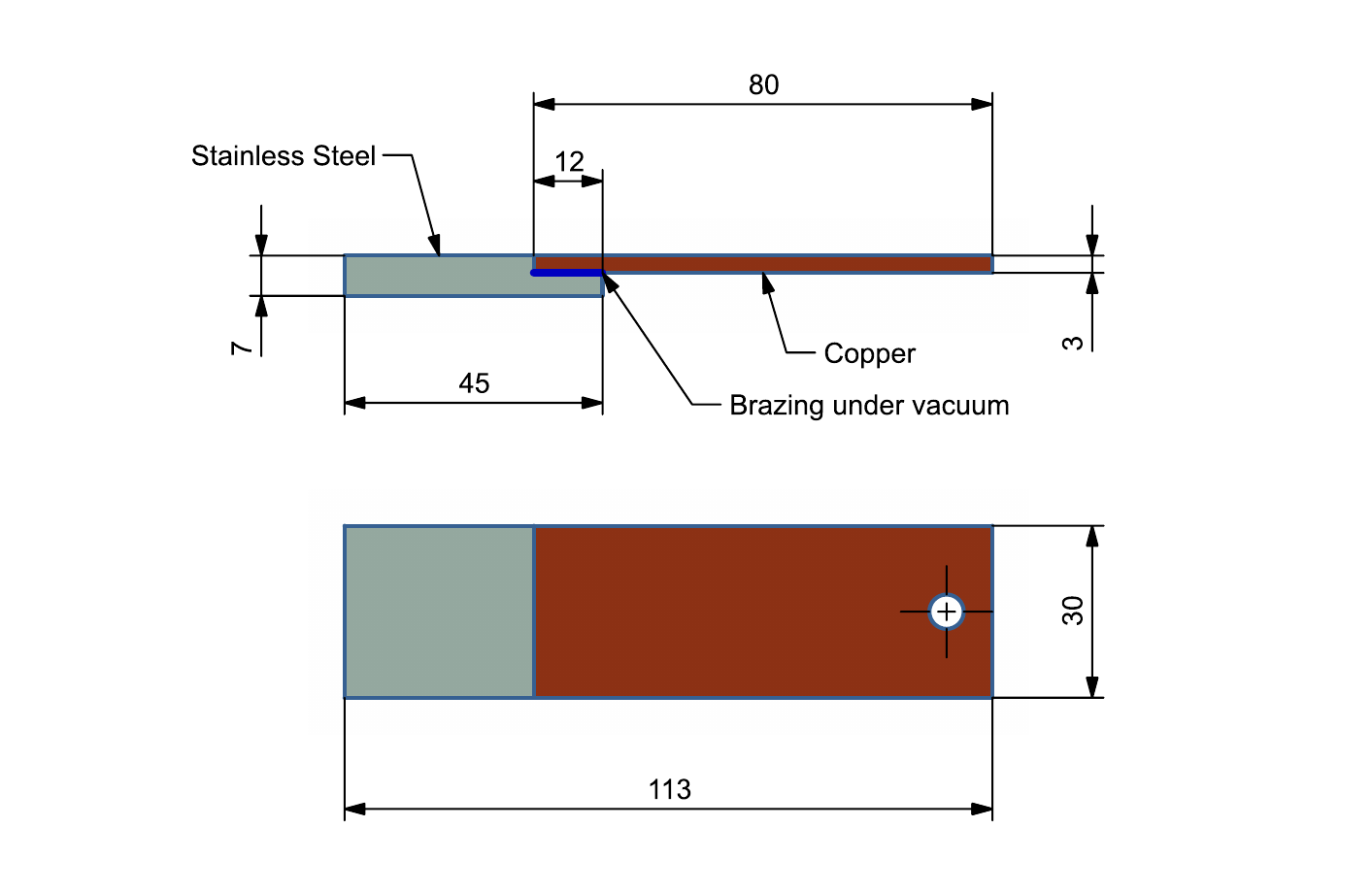}
\caption{Brazed sample drawing. The brazing paste area is highlighted in blue in the upper panel. Dimensions are in mm. 
\label{fig:dpb}}
\end{figure}

The brazed connection between OFHC copper and AISI 316 L steel cannot be simulated precisely by a finite element (FE) analysis, due to the unknown cryogenic properties of the brazing paste.
Therefore, to verify its robustness in operating conditions, a set of four brazed samples (Figure \ref{fig:dpb}) were prepared by the manufacturing company with the same two materials and brazing procedure employed for the final chamber. The samples underwent bending tests in LN$_2$, aiming to reproduce
the stress induced on the chamber brazed connection in operating conditions.

\begin{figure}
\centering
\includegraphics[width=\textwidth]{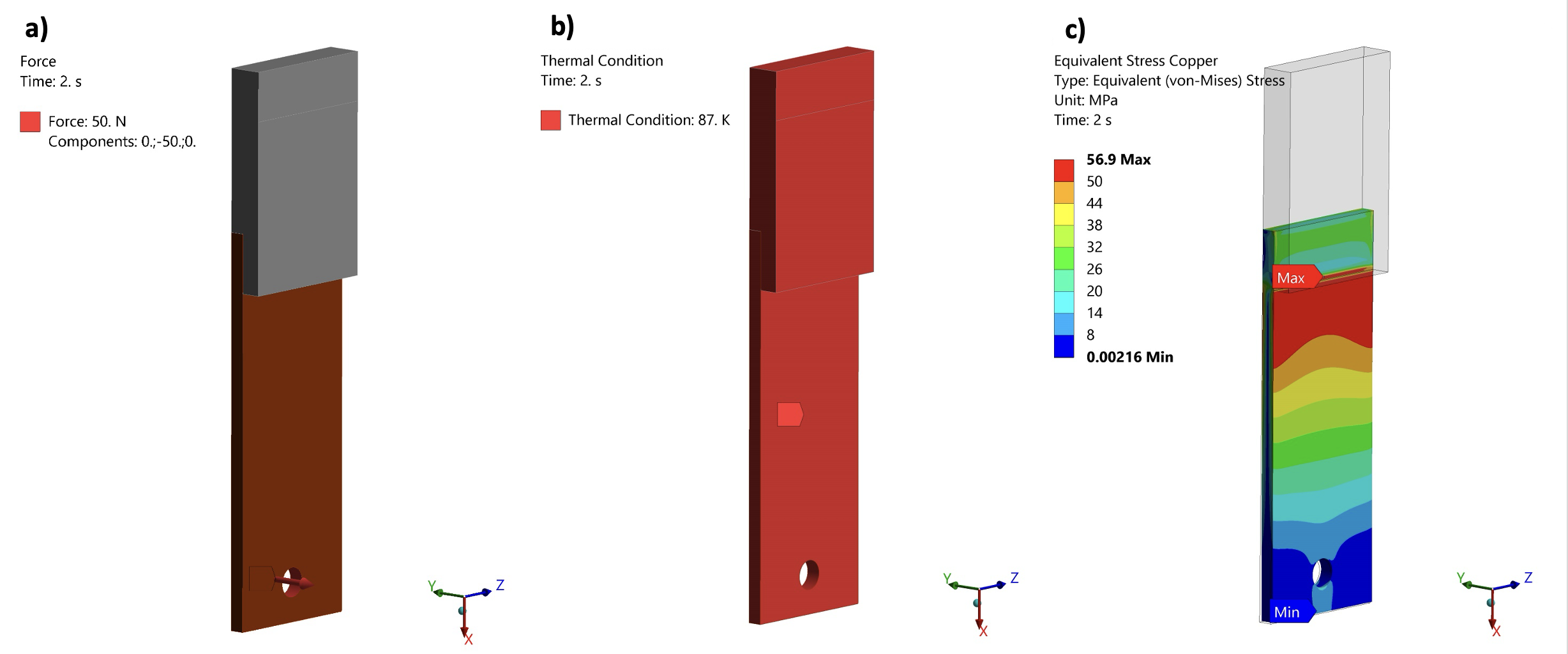}
\caption{Simulation of the brazed connection sample. Panel (a): variable normal force; panel (b): thermal load due to cooldown to 87~K; panel (c): Von Mises stress map obtained with a force of 50~N. 
\label{fig:spb}}
\end{figure}

A simulation of the samples was performed 
to determine the force to be exerted during the tests, corresponding to the maximum Von Mises stress of 57~MPa on OFHC copper in the brazed connection region, as indicated by the simulation of the full chamber (section \ref{sec:simulation_structural})\footnote{The value used here belongs to an earlier version of the chamber simulation and is 10\% higher than the final value reported in section \ref{sec:simulation_structural}. The test is therefore conservative.}.
The sample simulation is shown in Figure \ref{fig:spb}: panel (a) shows the variable normal force applied on the sample; panel (b) shows the thermal load expected during the test\footnote{The simulation assumed LAr temperature (87~K), although the tests were done in LN$_2$ (77~K) out of availability and are therefore conservative.}; the force was increased until obtaining the target stress shown in panel (c). 
The isotropic hardening of copper was included assuming a simplified bilinear curve. 
The simulation yielded a force of 50~N to be used in the experimental tests.

\begin{figure}
\centering
\includegraphics[width=.50\textwidth]{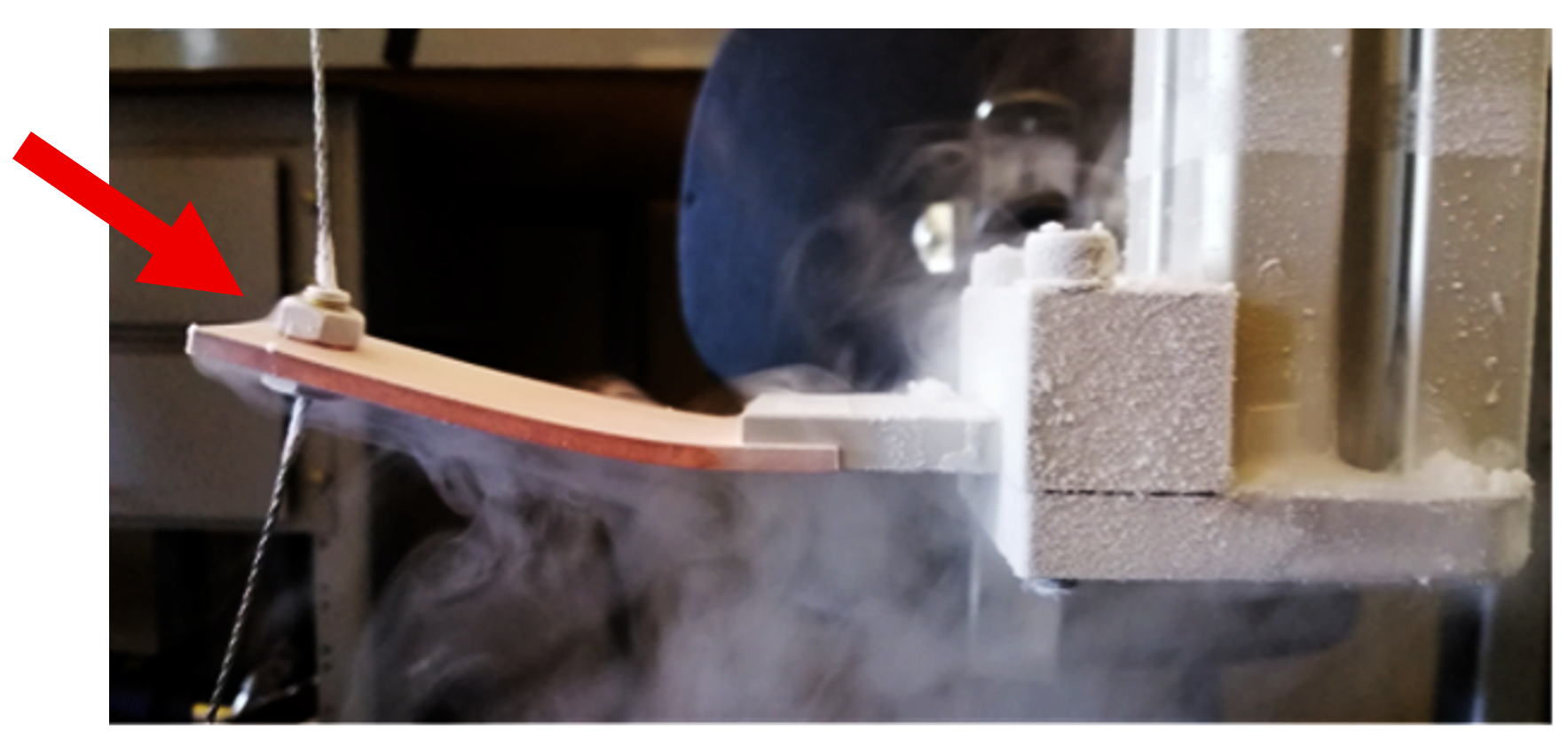}
\caption{Bent sample, suspended above the LN$_2$ bath after the test. The arrow highlights the connection point where the pulling force was exerted. 
\label{fig:fpb}}
\end{figure}

To carry out the tests a dedicated support structure was realized. The AISI 316 L end of the sample was clamped horizontally, whereas the OFHC copper end was left free (Figure~\ref{fig:fpb}). A stainless steel wire was connected to a hole at the OFHC copper end and to a dynamometer (Sauter FH100, 100 N capacity) fixed on a rail. By pulling the rail upwards, the dynamometer measured the normal force exerted on the sample. The sample was immersed in LN$_2$ for the duration of the test. Two samples were subject to the nominal force (50~N), whereas for two others a 1.5 factor was included, reaching 75~N.

A vertical deformation of about 8-10~mm and 16-17~mm was measured for the samples subject to 50~N and 75~N, respectively.
All four samples underwent X-ray scan before and after the tests, which showed no noticeable difference in the brazed connection, demonstrating the robustness under operating conditions for the actual chamber.

\section{Chamber simulation}
\label{sec:simulation}

The design of the cryogenic chamber described in Section \ref{sec:chamber} required validation under two respects. 
A thermal analysis was necessary to understand the value and uniformity of the temperature field in the inner volume, given the power dissipated through the heater. 
A structural analysis was necessary to understand if the stress expected in the most challenging operative conditions would pose safety risks or induce excessive deformations.

To this aims, two coupled multiphysics analyses were carried out, in parallel and independently by the Mechanical and Design Services of INFN Milan and LNGS. For the thermal analysis, the first simulation focused on a solid model, by means of a FE analysis (INFN Milan and LNGS), while the second considered a fluid model, by means of a computational fluid dynamics (CFD) simulation (INFN LNGS). In each case, the temperature field issued by the simulations was used as thermal boundary condition for the subsequent structural chamber evaluation. We present here a summary of the results, while complete reports of these works are available in~\cite{milano2,lngs2023}.

\subsection{Comparison between CFD and FE approach}
\label{sec:simulation_comparison}

CFD, based on the Navier-Stokes equations (i.e., conservation of mass, momentum, and energy), can simulate the behavior of a gas by considering the contribution of conduction, convection, and radiation, and allows for the description of turbulence within the fluid. FE, by contrast, models heat transfer in the gas volume by conduction only, neglecting other contributions; for this reason, it is computationally simpler to implement and run, and it requires fewer physical parameters. 

The boundary and operating conditions of the ASTAROTH cryogenic chamber indicate that heat exchange between fluids is slow and primarily governed by conduction. Consequently, no substantial difference is expected between FE and CFD models. 
However, the presence of a high thermal gradient between the copper-steel double wall suggests a locally more complex gas behavior, which could be better described by a fluid model.

\begin{figure}
\centering
\includegraphics[width=.70\textwidth]{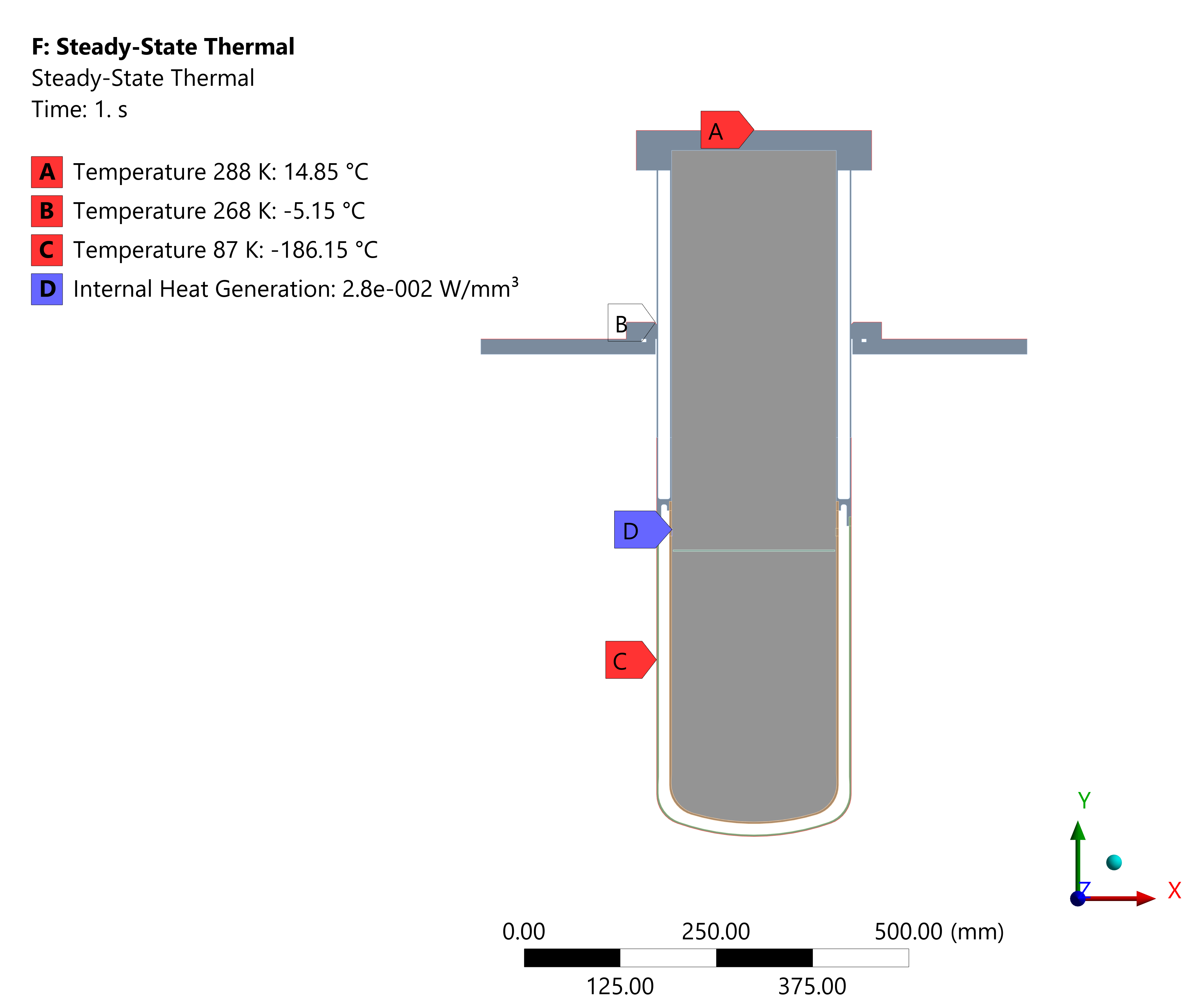}
\caption{Two-dimensional (2D) model of the cryogenic chamber and boundary conditions applied.
\label{fig:2D}}
\end{figure}

For this reason, before carrying out a 3D thermo-structural analysis including all the complexity of the chamber, a comparison between the two approaches was performed based on the simplified 2D model shown in Figure~\ref{fig:2D}. The model includes the double-walled copper-steel chamber and chimney, the thermal bridge, the lowest anti-convection disk, the top flange and the outer dewar flange. The following boundary conditions have been considered: the top flange was set at 288~K, the dewar flange at 268~K\footnote{This temperature is an estimate based on our previous experience with boiling liquid nitrogen dewars.} and the outer wall of the chamber in contact with the LAr at 87~K. The heater was assumed to dissipate 190~W (corresponding to $2.8\times 10^{-2}$ W/mm$^3$ in the copper). This value was analytically calculated to be the power required to maintain the inner volume at the upper limit of the design range (150~K), which also represents the most demanding operational condition for the materials.  The software used for the simulations is ANSYS Workbench: the “Fluent” environment for the CFD model and the “Mechanical” one for the FE model.

\begin{figure}
\centering
\includegraphics[width=.85\textwidth]{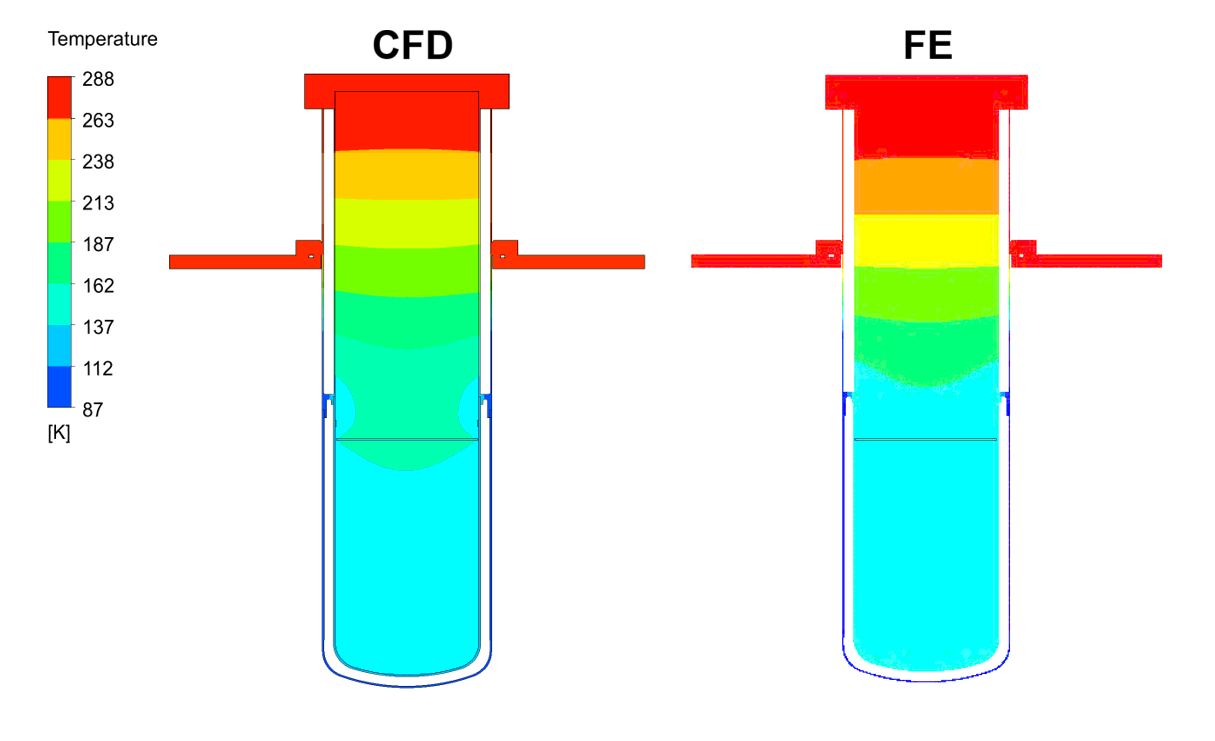}
\caption{CFD and FE model: resulting thermal field within the He volume. \label{fig:cfd_vs_fem}}
\end{figure}

The thermal fields within the He volume resulting from the two approaches are shown in Figure~\ref{fig:cfd_vs_fem}. 
The comparison shows that the two methods yield comparable results across most of the volume.  
However, the CFD model, despite some simplifications (e.g., assuming laminar flow), appears to capture the critical area around the thermal bridge and the copper-steel connection in more details. 

\begin{figure}
\centering
\includegraphics[width=.80\textwidth]{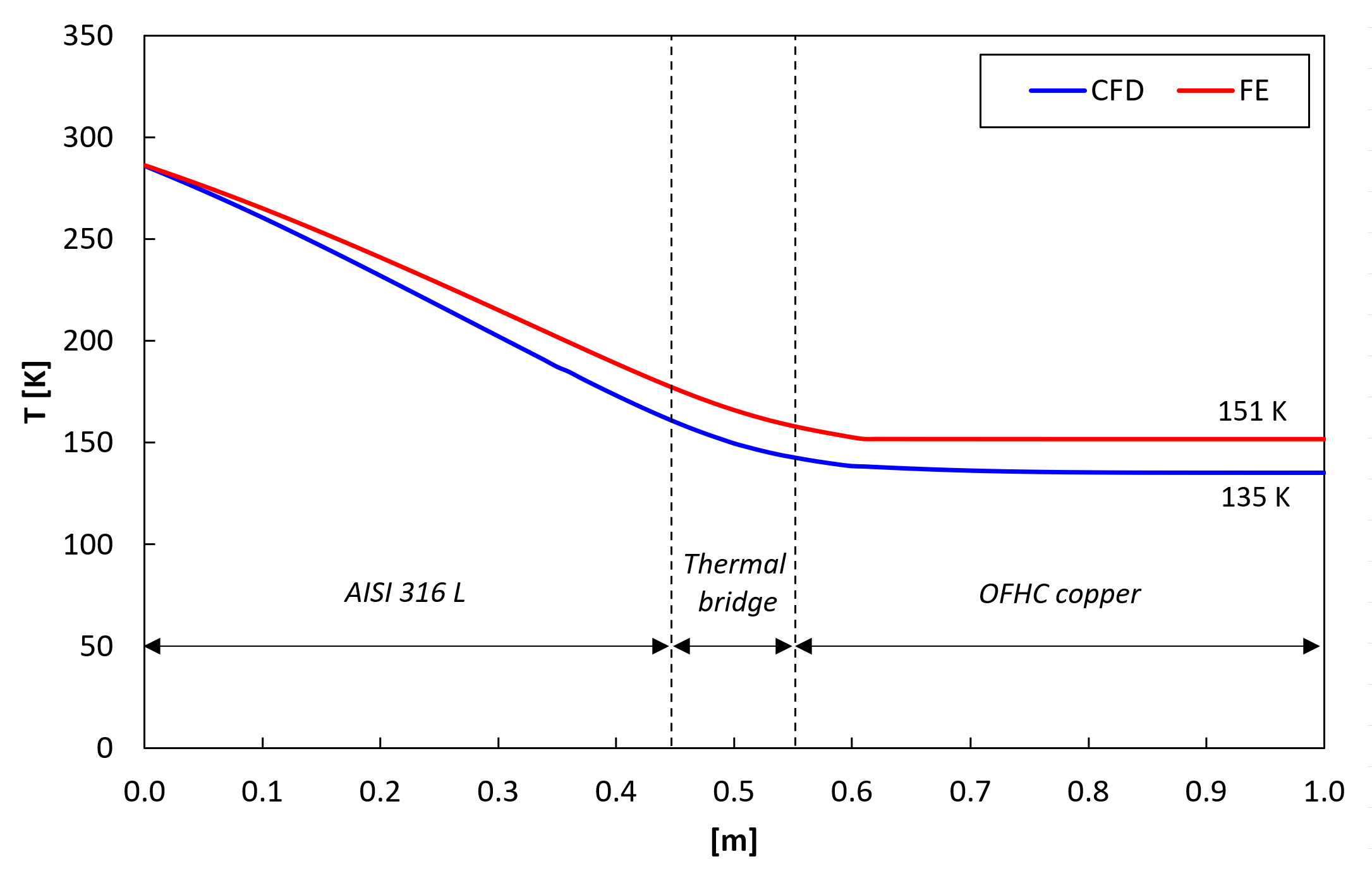}
\caption{CFD and FE model: temperature vs distance from the top flange along the symmetry axis of the chimney and chamber construct.  \label{fig:T_axis}}
\end{figure}

To quantify the difference between the two models, the temperature  along the axis of symmetry of the cryogenic chamber is shown in Figure~\ref{fig:T_axis}, starting from the top flange. A small discrepancy begins to appear near the thermal bridge and affects mostly the lower part of the volume, where a maximum difference of $\sim 15$~K can be observed.

It was decided to proceed to the 3D thermal simulation using the FE approach, given the good agreement with the target temperature of 150~K and after considering computational efficiency\footnote{FE Method is not only simpler and faster, but it allows for automatic mapping of the thermal field for the subsequent structural analysis based on the same method.}.

\subsection{Thermal analysis}
\label{sec:simulation_thermal}

\begin{figure}
\centering
\includegraphics[width=0.9\textwidth]{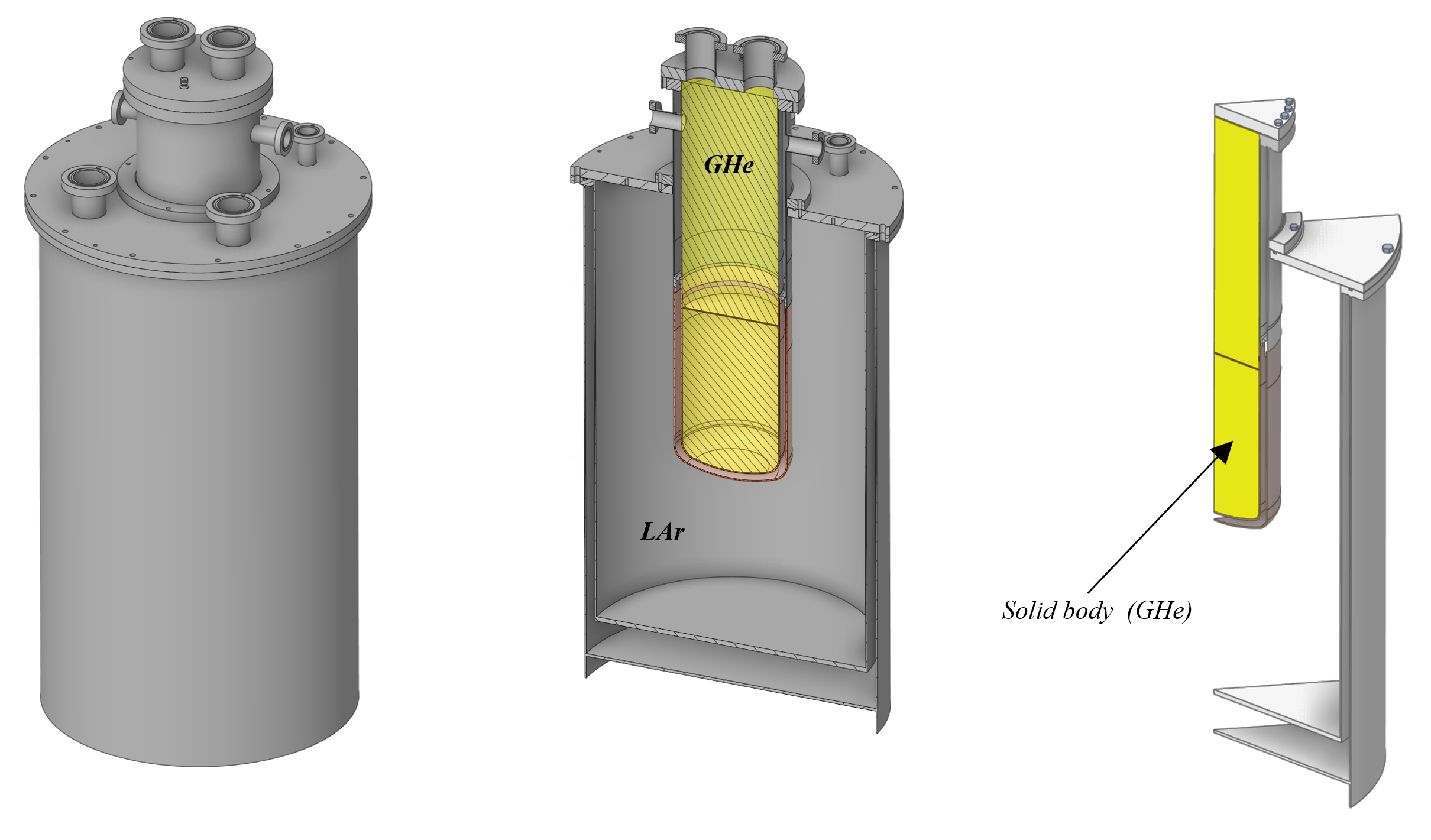}
\caption{Three-dimensional (3D) model of the chamber. 
\label{fig:3D}}
\end{figure}

The 3D multiphysics analysis of the cryogenic chamber and its dewar was thus carried out using the ANSYS Workbench software (Mechanical environment), modeling the He volume as a solid in which heat is transferred solely by conduction.
The mesh size was set to 5~mm for the dewar and 3~mm for the chamber and chimney components. 
The chamber model was defeatured by removing details not relevant to the analysis, such as the UHV ports, fillets, chamfers, etc. The system cylindrical symmetry was exploited to simulate only 1/8 of the full model (Figure~\ref{fig:3D}).
This simplification reduced the computational time and allowed for the refinement of the mesh to 1.5~mm in the critical area of the thermal bridge and copper-steel junction. The thermal analysis was performed using the same boundary conditions described in the previous section.

\begin{figure}
\centering
\includegraphics[width=0.8\textwidth]{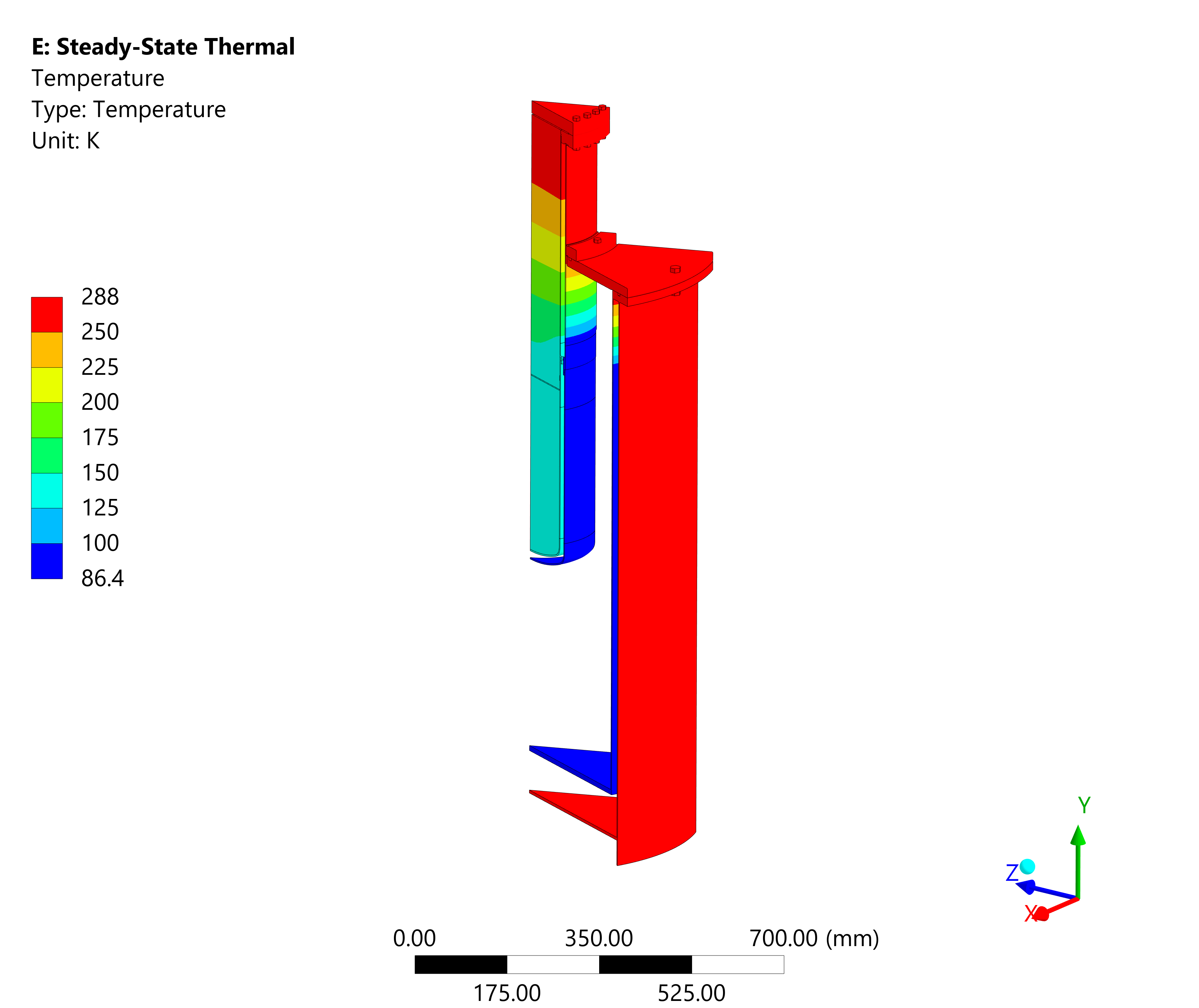}
\caption{Thermal field resulting from the 3D thermal analysis. 
\label{fig:thermal_field}}
\end{figure}
The resulting thermal field is shown in Figure~\ref{fig:thermal_field}. The temperature distribution confirms the findings of the preliminary 2D analysis.
This thermal field was then used as an input boundary condition for the structural analysis of the chamber.

\subsection{Structural analysis}
\label{sec:simulation_structural}

For the structural analysis, the elasto-plastic model of OFHC copper, as described in Section \ref{sec:materials_mechanical}, has been used within the finite element analysis to describe the temperature-dependent properties of the material.
In addition to the temperature field obtained from the thermal analysis, the following boundary conditions were applied (Figure~\ref{fig:boundary}):
\begin{figure}
\centering
\includegraphics[width=0.8\textwidth]{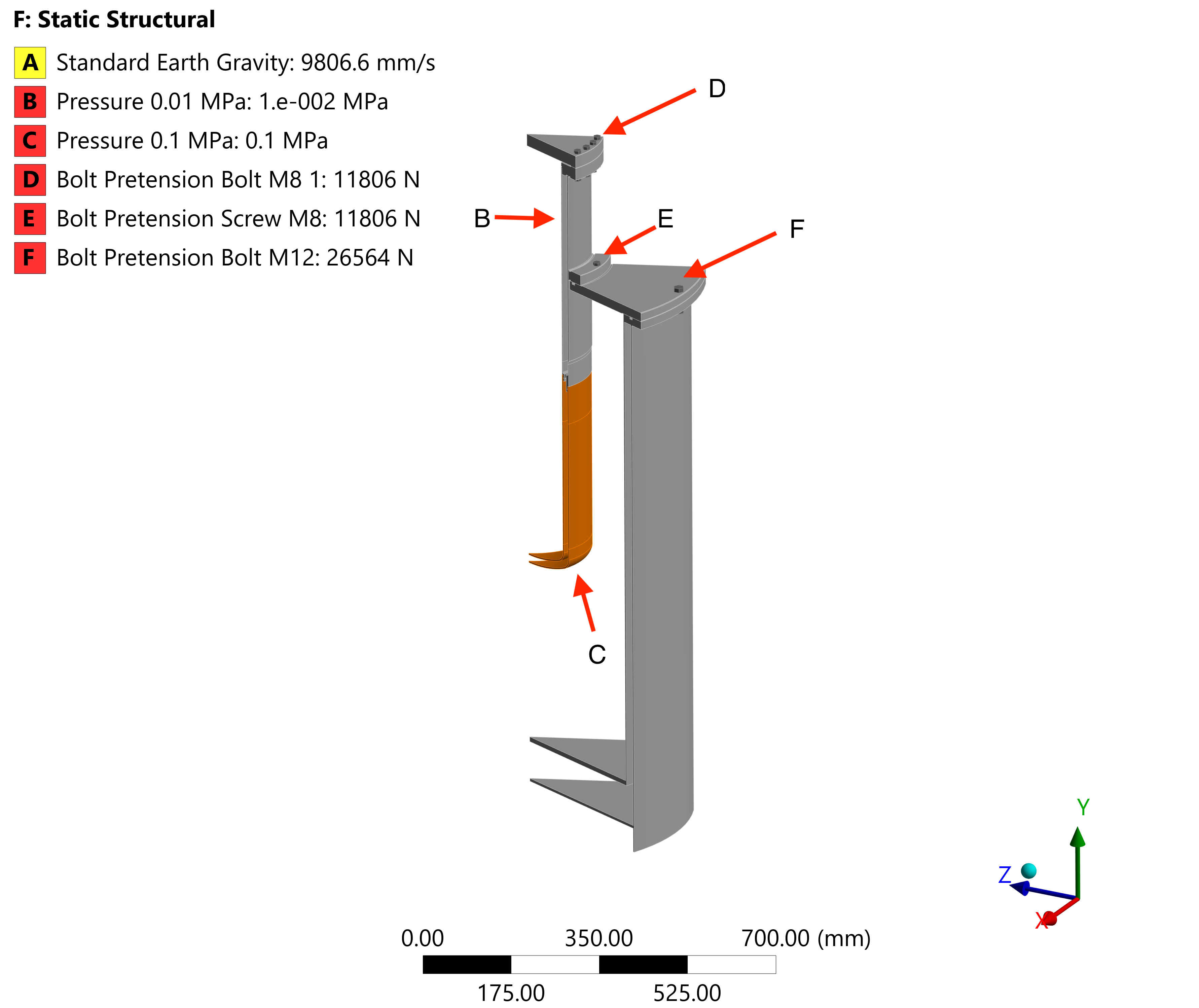}
\caption{Boundary conditions of the 3D structural analysis. 
\label{fig:boundary}}
\end{figure}

\begin{itemize}
    \item A pressure of 0.1~MPa was applied to the outer wall in contact with LAr, neglecting variations along the chamber height, which are $\leq 70$~mbar.\footnote{Preliminary simulations showed that mechanical loads due to pressure in the range 100--1000~mbar are negligible compared to thermally induced loads.}
    \item A pressure of 0.01~MPa was applied to the inner wall in contact with GHe.
    \item The pre-load of the bolts was set to one-fourth of their maximum load.
    \item A static friction coefficient of 0.8 was assumed between the steel flanges.
    \item The effect of gravity was included.
    \item The mechanical response of the ground supporting the dewar was included.
\end{itemize}

\begin{figure}
\centering
\includegraphics[width=0.85\textwidth]{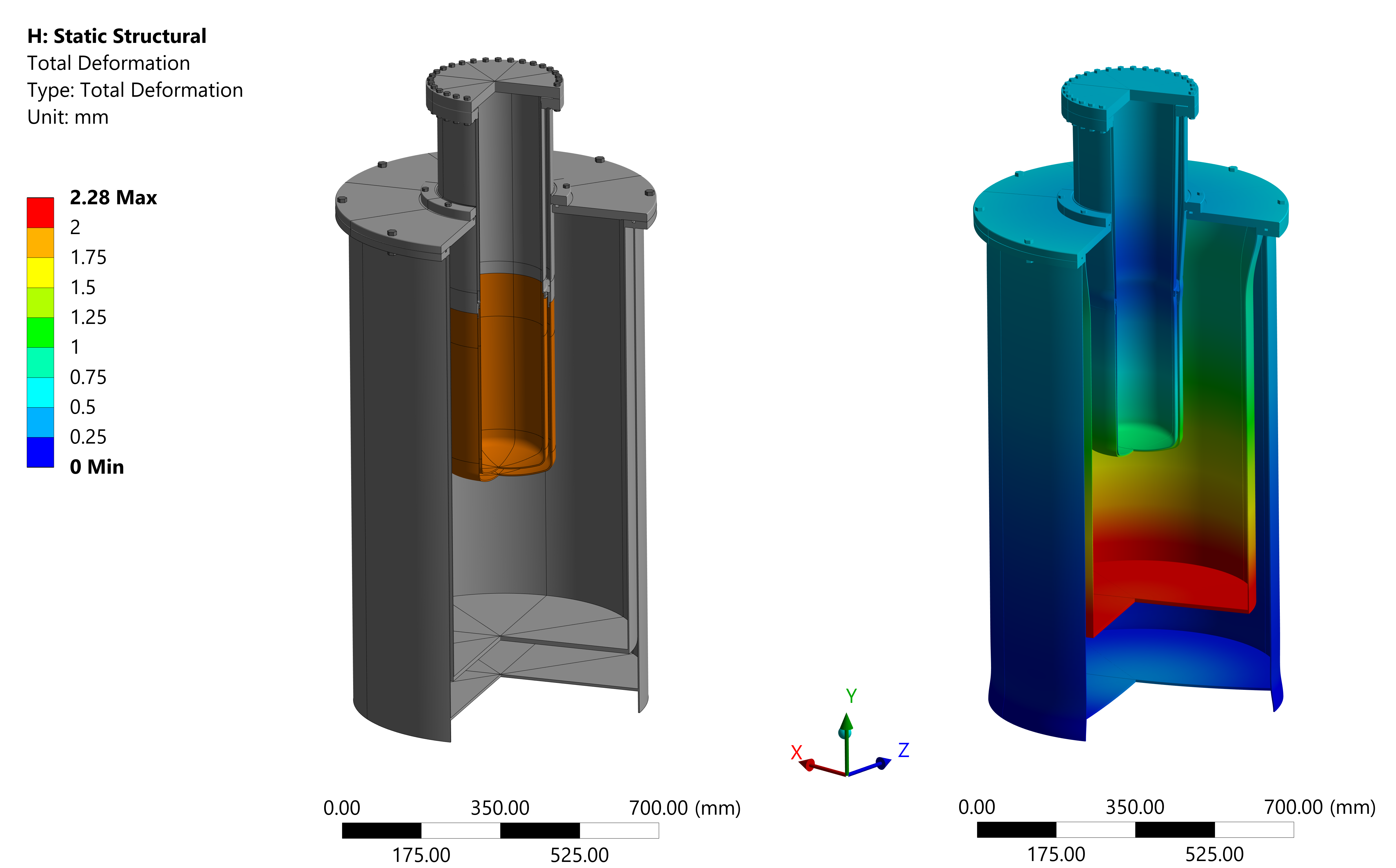}
\caption{The right image shows the deformation magnified by a factor 40 and the actual deformation field  in color scale. The left image indicates copper (orange) and steel (gray) sections for reference.}
\label{fig:results_chamber}
\end{figure}

\begin{figure}
\centering
\includegraphics[width=1.0\textwidth]{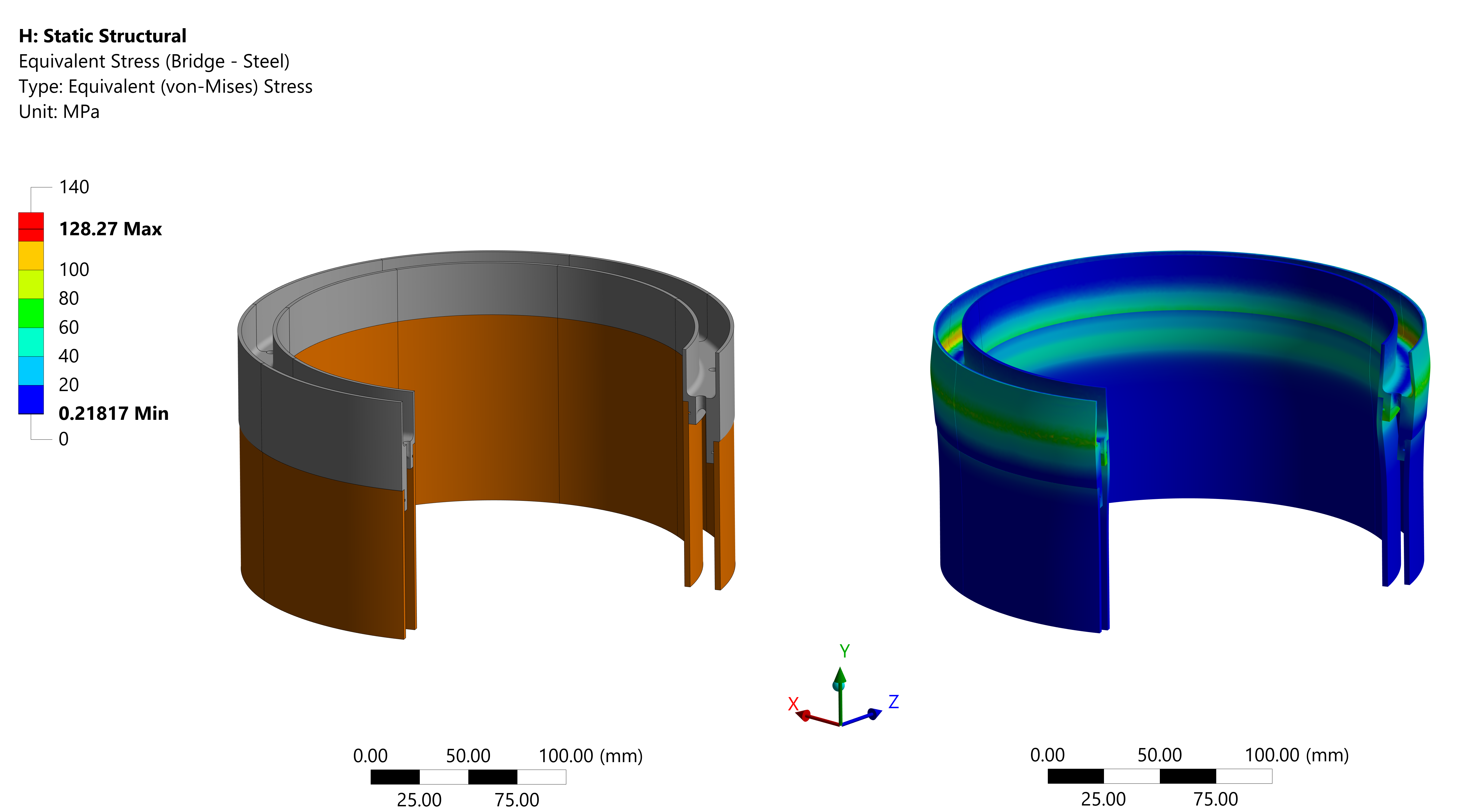}
\caption{Stress distribution near the thermal bridge (copper-steel junction), shown in color scale on the right. The deformation is magnified by a factor 50. The left image indicates copper (orange) and steel (gray) sections.}
\label{fig:results_bridge}
\end{figure}

\begin{figure}
\centering
\includegraphics[width=0.70\textwidth]{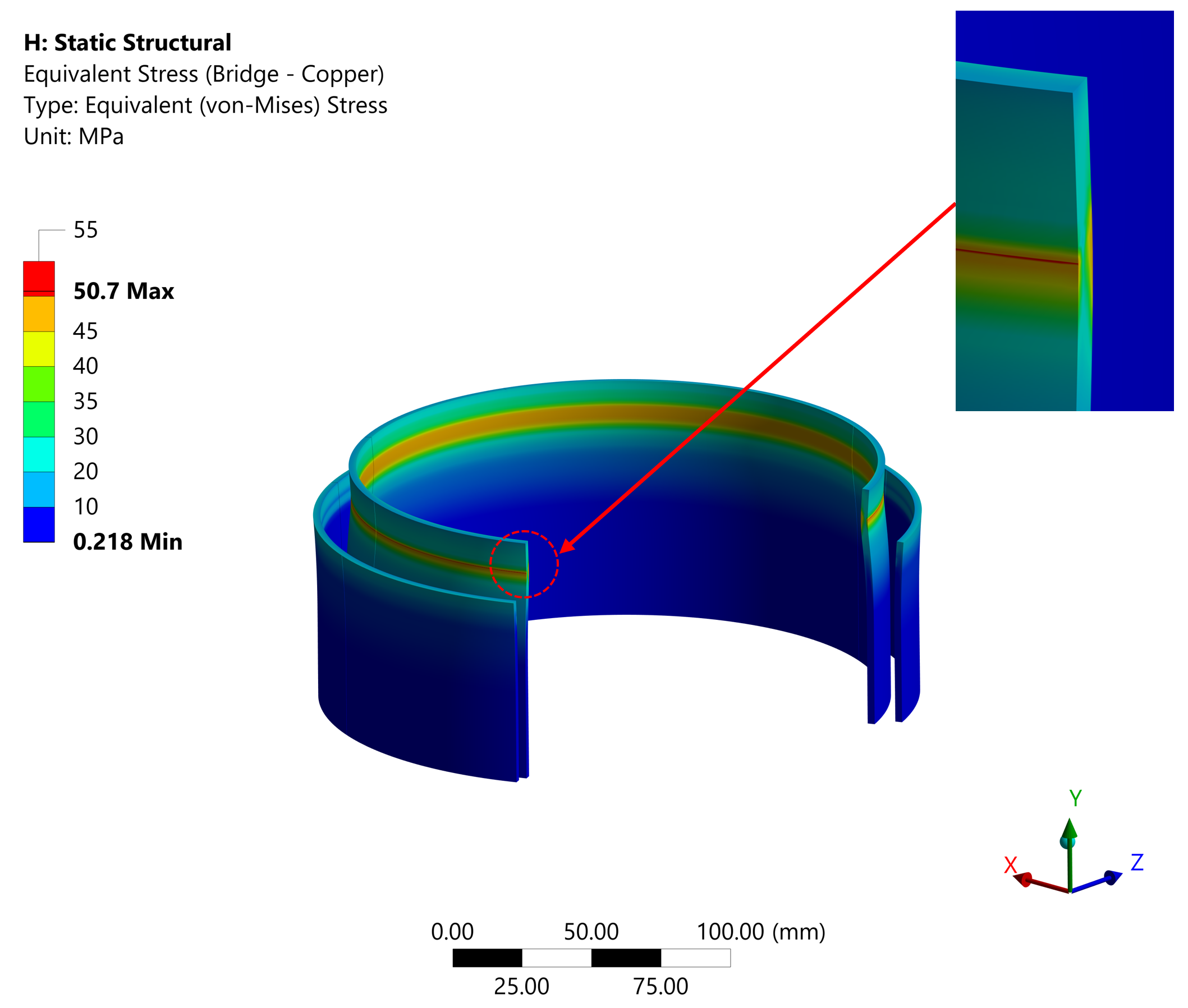}
\caption{Stress distribution in the copper wall immediately below the copper-steel junction. The deformation is magnified by a factor 50.
\label{fig:results_junction}}
\end{figure}

Figure~\ref{fig:results_chamber} shows the deformation field resulting from the structural simulation, operating at around 140~K. 
The maximum deformation occurs at the bottom of the external dewar, reaching approximately 2.3~mm. In the OFHC copper chamber, deformation remains below 1.0~mm. Figure~\ref{fig:results_bridge} displays the stress distribution near the thermal bridge and copper-steel junction. The highest stress, about 130~MPa, appears on the outer steel wall: the value is well below the YS of the material at all relevant temperatures (see Table~\ref{tab:316-mech}). The maximum stress in copper is found on the inner wall, immediately below the brazed connection. The value of $\sim$~51~MPa exceeds the YS limit at 140~K, $\sim$~36~MPa, estimated by linear interpolation from the values in Table~\ref{tab:copper-mech}. This indicates that copper in this region is, as expected, in a strain-hardened condition. The highest stress value is however well below the UTS limit, guaranteeing the structural integrity of the chamber. Moreover, a finer analysis of the stress field in this region (Figure~\ref{fig:results_junction}) shows that stresses above the YS limit are only localized in a surface band. In the core of the material, instead, stress values are lower, mostly ranging from 15 to 30~MPa. 

During commissioning, after a slow cooling to 87~K, the temperature should be gradually raised to 150~K through a sequence of heating and cooling cycles. This procedure is designed to induce strain hardening in the OFHC copper, thereby increasing its effective YS and the overall safety. Following this treatment, the system is expected to operate within the elastic regime across the entire temperature range. Eventually, the total number of thermal cycles over the system lifetime is not expected to exceed a few tens, making considerations related to low cycle fatigue unnecessary.

\section{Chamber commissioning and operation}
\label{sec:commissioning}

\begin{figure}
\centering
\includegraphics[width=.40\textwidth]{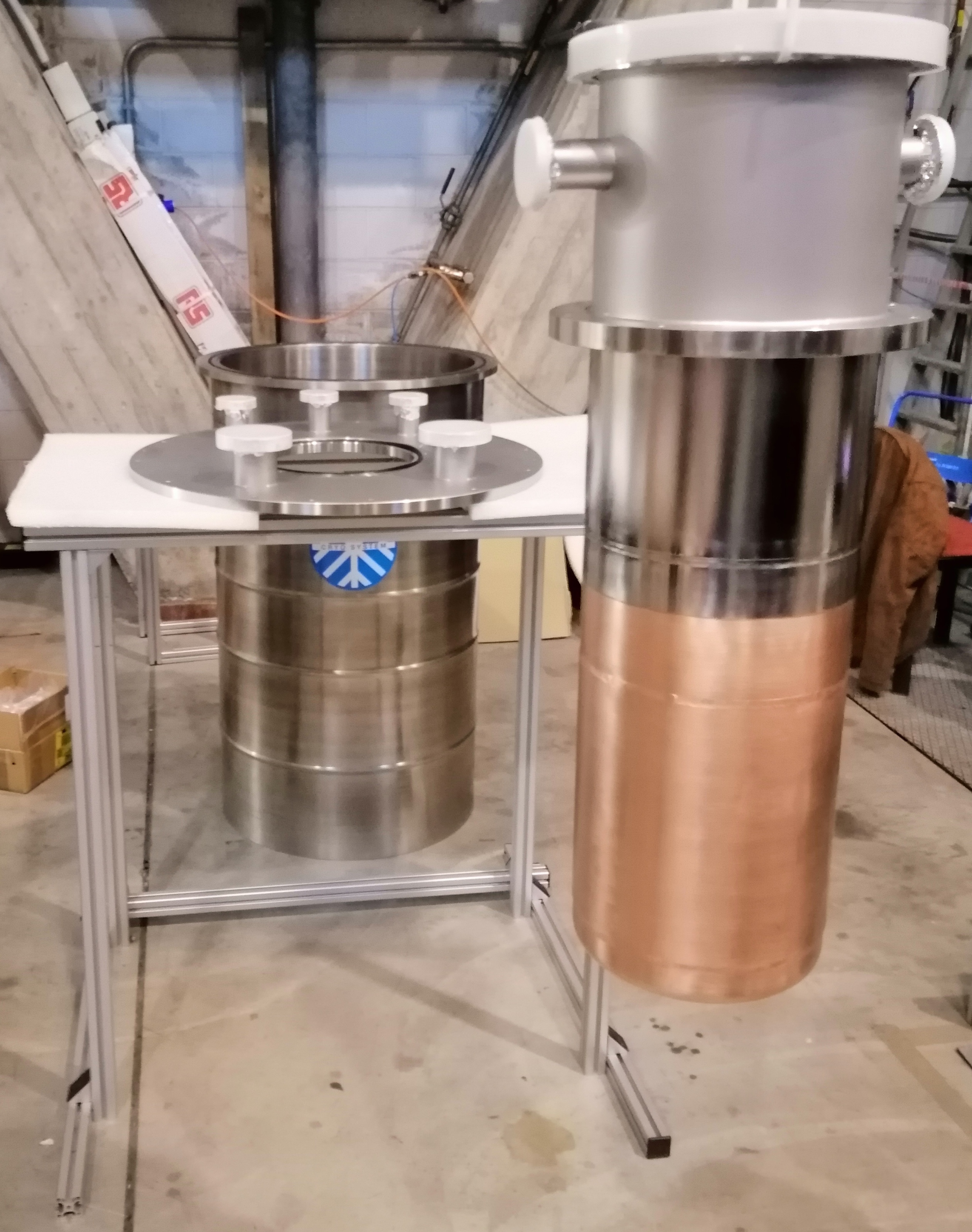}
\caption{Chamber delivered at LASA, before assembly into the external dewar. \label{fig:foto_criostato}}
\end{figure}

After the design and validation process, the chamber was constructed and assembled into the external dewar at LASA as shown in Figure~\ref{fig:foto_criostato}. The system was commissioned with a first run without crystals in September 2023, following the recommendations outlined in Section~\ref{sec:simulation_structural}. Since then, the cryostat has been in regular operation. 
In late 2023, the first four cooling campaigns focused on the characterization of a test ASIC~\cite{andreani} across the operational temperature range, validating the ASIC technology for cryogenic digital applications\footnote{Although this study is of general interest, it was performed within ASTAROTH in order to select the ASIC technology for the upcoming SiPM front end with charge mapping capability, that is a key development of the project.}. 
Starting in 2024, several runs with a NaI(Tl) crystal were performed, investigating several SiPM matrices and front-end electronics solutions. Some results from these campaigns are reported in~\cite{Martinenghi2025}.

\subsection{Differences between model and operation}

There are two differences between the design parameters used in the simulation and the actual operativity of the chamber in this early phase.

(1) The designed cryogen assumed in the model is LAr due to the possibility to exploit the outer volume as a veto detector, as explained in section \ref{sec:astaroth}. However, in all runs performed to date, LN$_2$ was more readily available and was used.  With this cryogen the theoretical minimum chamber temperature is 77~K, neglecting losses. As soon as the outer volume will be instrumented with light sensors, LAr will be used.

(2) The design pressure for the helium volume is 100~mbar to guarantee the slow and safe thermalization of the detectors. However, the current crystal case, made of fused silica, encloses the NaI crystal with a 1~mm neon gas gap at 1~bar to compensate for the different coefficients of thermal expansion of the two materials.
Because this design cannot safely withstand a positive pressure differential, the inner chamber volume has so far been filled at 1~bar, kept constant throughout the thermal cycles.
Since the present encasing is suboptimal for light collection~\cite{Martinenghi2025}, a new design has been finalized, in which the crystals are coated with epoxy resin.
For these detectors, the chamber helium atmosphere will be set at the design pressure of 100~mbar.

\subsection{Performance}

To monitor operations and performance, the cryostat is equipped with the following sensors:
\begin{itemize}
    \item Three RTD PT100 sensors installed on a support bar in the outer volume, located at: 5~cm from the top main flange; halfway down the cryostat; and 5~cm from the bottom. These are used to estimate the LN$_2$ level inside the dewar.
    \item One RTD PT100 sensor installed 22~cm from the chamber bottom on the inner copper wall.
    \item A manometer and a relative pressure sensor (max $+0.25$~bar, accuracy $\pm0.25\%$ full scale, 4–20~mA) connected to the inner chamber to monitor and stabilize the He gas pressure during cool-down and warm-up phases.
\end{itemize}

Due to minor heat losses, the lowest temperature reached was 82~K, a few hours after completing the LN$_2$ filling. A temperature stability of 0.1~K has been consistently demonstrated across different setpoints, over time frames of a few hours. The upper end of the temperature range (150~K) can be maintained by injecting a power of 177~W. This value is somewhat lower than the value computed analytically (190~W) and used for the simulations ( Section~\ref{sec:simulation_comparison}). The difference is not unexpected considering that the computation was assuming the design 100~mbar gas pressure, while the chamber was operated at 1~bar. 

\subsection{Future developments}

At present, LN$_2$ is left to evaporate during data taking, requiring periodic refilling. Consequently, the heater power must be manually adjusted to keep the temperature stable throughout the cycles and irrespectively of the dewar filling level.

Future developments of the system include: (1) automatic tuning of the heating power, by means of a PID controller, to further improve temperature stability; (2) instrumenting the outer volume with SiPMs and switching to liquid argon to serve as a veto detector; and (3) implementation of a closed-loop argon recirculation and purification system, eliminating the need for refills and minimizing cryogen consumption.

\vspace{5mm}

To date, the chamber has undergone about 30 cooling cycles and multi-day runs without any degradation in performance. This successful operation demonstrates both the strength and effectiveness of the design and the validity of the simulations.

\section{Conclusions}
\label{sec:conclusions}

ASTAROTH is a project aimed at the development of next-generation NaI(Tl)-based detectors for the direct search for dark matter.
It employs SiPMs, which, for low-energy applications, must be operated at cryogenic temperatures in order to strongly suppress the dark count rate and take advantage of their several benefits over traditional PMTs.

This document presents the innovative design of a cryostat based on a double-walled, vacuum-insulated OFHC copper chamber. The chamber is immersed in a cryogenic bath and hosts the detectors in a helium atmosphere, ensuring uniform thermalization. The design balances the cooling power of the cryogenic bath reaching the inner volume through a specially designed steel bridge element, and the power dissipated by a heater.
As a result, the internal temperature can be tuned at any value in the 87–150 K range.

We have discussed the experimental characterization of the thermo-mechanical properties of the materials used, at both room and cryogenic temperatures. The simulations developed to validate the design, including thermal and structural analysis, have also been described. In particular, thermal analysis compared FE and CFD methods to a simplified 2D model, showing that the design goals are met and that there is substantial agreement between methods. This is consistent with the fact that heat transfer in the system occurs primarily through conduction.
A full 3D FE simulation was then performed, producing a more accurate temperature distribution. This thermal map was used as the input for the subsequent structural analysis. 
 
Results indicate that, near the thermal bridge (copper-steel junction), the stress marginally exceeds the YS of copper at the maximum operating temperature. However, this condition is limited to a narrow surface band and does not compromise the chamber mechanical integrity. Moreover, controlled cycling during commissioning is expected to induce strain hardening in this region, locally raising the YS limit and leading to subsequent operation in the elastic regime.

The chamber was commissioned with LN$_2$ and is now in regular operation. In the next phase of the project, the outer volume will be instrumented with additional SiPMs, and the system will switch to liquid argon as foreseen in the baseline design. This will allow the outer volume to act as a veto detector against gamma-accompanied backgrounds.

The chamber has now undergone around 30 cooling cycles without signs of performance degradation. The internal temperature has been successfully stabilized within 0.1~K at arbitrary values in the 82–150~K range. Thanks to this successful design, a rich program of characterization of electronic components, SiPM arrays and crystals for the development of ASTAROTH is ongoing since October 2023.

The cryostat performance and robustness prove its suitability not only for the ASTAROTH detector, but also for a wide range of applications requiring uniform, tunable gas-conducted cryogenic environments.

\acknowledgments

This work has been supported by the Fifth Scientific Commission (CSN5) of the Italian National Institute for Nuclear Physics (INFN), through the ASTAROTH grant.

Authors would like to acknowledge the support received from the staff of the “Accelerators and Applied Superconductivity Laboratory” (LASA) in Milan. In particular the work of Maurizio Todero and Arsenio Palmisano has allowed the characterization of the OFHC copper samples, while Danilo Pedrini and Augusto Leone have been fundamental in operating the system.

\bibliographystyle{JHEP}
\bibliography{biblio.bib}

\end{document}